\documentclass[twocolumn]{aastex62}
\usepackage{multirow}
\usepackage{comment}
\usepackage{hyperref}
\usepackage{url}
\usepackage{booktabs}
\usepackage[caption = false]{subfig}
\usepackage{natbib}

\def\degs{\ifmmode ^{\circ}\else$^{\circ}$\fi}
\def\fdg{\hbox{$.\!\!^\circ$}}          

\submitjournal{ApJ}

\shorttitle{BALROG verification}
\shortauthors{Berlato et al.}

\begin{document}
	
	\title{Improved Fermi-GBM GRB localizations using BALROG}
	
	\author{F. Berlato}
        \affiliation{Max-Planck Institut f\"ur extraterrestrische Physik, 85748 Garching, Giessenbachstr. 1, Germany}
        \affiliation{Physics Department, Technische Universit\"at M\"unchen, 85748 Garching, James-Franck-Strasse 1, Germany}

	\author{J. Greiner}
        \affiliation{Max-Planck Institut f\"ur extraterrestrische Physik, 85748 Garching, Giessenbachstr. 1, Germany}
        \affiliation{Excellence Cluster Universe, Technische Universit\"at M\"unchen, Boltzmannstr. 2, 85748, Garching, Germany}
	
	\author{J. Michael Burgess}
		\affiliation{Max-Planck Institut f\"ur extraterrestrische Physik, 85748 Garching, Giessenbachstr. 1, Germany}
		\affiliation{Excellence Cluster Universe, Technische Universit\"at M\"unchen, Boltzmannstr. 2, 85748, Garching, Germany}
	
	\correspondingauthor{Francesco Berlato}
	\email{fberlato@mpe.mpg.de}
	
	\begin{abstract}
          The localizations of gamma-ray bursts (GRBs) detected with the
          Gamma-ray Burst Monitor (GBM) onboard the Fermi satellite
          are known to be affected by significant systematic errors of
          3--15 degrees.  This is primarily due to mismatch of the
          employed Band function templates and the actual GRB
          spectrum.  This problem can be avoided by simultaneously
          fitting for the location and the spectrum of a GRB, as
          demonstrated with an advanced localization code, BALROG \added{\citep{2018MNRAS.476.1427B}}. Here, we
          analyze in a systematic way a sample of 105 \added{bright} GBM-detected GRBs for which accurate reference localizations
          are available from the {\it Swift} observatory. We show that
          \replaced{the previous systematic errors are for the vast majority generated by the use of spectral templates and that GBM's GRB localizations can be
            substantially improved by allowing for the simultaneous
            fit of both position and spectrum.}{the remaining systematic error can be reduced to $\sim$1--2\degr.}
	\end{abstract}
	

	\section{Introduction}
	
	Gamma-ray bursts (GRBs) are the most energetic electromagnetic
        phenomenon in the Universe, releasing up to
        $E_{iso}\sim 10^{54}$ erg. They consist of bright flashes of
        gamma-ray photons, which can last from a fraction of second to
        a few thousand of seconds.  The initial gamma-ray emission
        (so-called prompt emission) is followed by a longer lasting,
        low-energy emission (so-called afterglow), ranging from X-rays
        to radio. While dependent on the wavelength and the specific
        burst considered, from a practical standpoint afterglows can
        typically be observed only for the first day after the prompt
        emission, before they become too faint for most instruments.
        The afterglow provides a wealth of information on the GRB's
        local environment and host galaxy, which is of great
        importance to improve our understanding of the GRB
        progenitor. An afterglow detection requires a good gamma-ray
        location since most follow-up telescopes have a narrow
        field-of-view which makes it unfeasible to survey large
        regions of the sky before the afterglow fades.

	Well-constrained localizations of the prompt gamma-ray
        emission are thus fundamental, not only for multi-wavelength
        observations of GRBs, but also for multi-messenger
        astronomy. The recent identifications of a gravitational wave
        \added{\citep{2017ApJ...848L..13A, 2017PhRvL.119p1101A, 2017ApJ...848L..14G}} counterpart for gamma-ray sources
        clearly demonstrates the importance of providing reliable
        locations. Since the probability of successfully identifying a
        counterpart in other wavelengths (or messengers) is inversely
        proportional to the size of the error region, any improvement
        in localization accuracy is important.
	
	Current Fermi-GBM locations of GRBs are known to be affected
        by large systematics \citep{2015ApJS..216...32C}.
        \replaced{Consequently, optical counterpart searches in GBM error boxes
        of GRBs had only 8/35 successful identifications by iPTF
        \citep{2015ApJ...806...52S}}{This represents a major obstacle for successful GRB afterglow identification \citep{2015ApJ...806...52S}}. Yet, Fermi-GBM is the most prolific
        GRB detector presently operational, triggering on about 270
        GRBs/yr, as well as hundreds of other transient source types
        like solar flares, Soft Gamma Repeaters, or Terrestrial
        Gamma-ray Flashes.
	
	Recently, it was shown that the large systematic error in
        Fermi-GBM localizations of GRBs is due to the spectral
        template deviating from the actual GRB spectrum, and can be
        eliminated when simultaneously fitting for the GRB position
        and spectrum \citep{2018MNRAS.476.1427B}. This was exemplarily
        shown for a handful of GRBs\added{, including bright as well as faint GRBs.}
	
	Here, we analyze in a systematic way a larger sample of GRBs which had been
	detected by both the Fermi-GBM detector and localized by the
	Neil Gehrels {\it Swift} observatory, thus providing accurate reference 
	positions. After an introduction of the GBM detector (\S 2), the 
	two different localization methods (\S 3) and the sample selection (\S 4),
	\S 5 presents the new localization results, \S 6 analyzes remaining
	deficiencies in the new localizations \added{and \S 7 suggests guidelines for best localization results using BALROG.}

	\section{The Gamma-ray Burst Monitor}

	The Gamma-ray Burst Monitor (GBM) on board the Fermi space
        telescope is an array of detectors built with the specific
        purpose of observing gamma-ray bursts. GBM is operating since
        2008 and as of now has triggered on more than 2300 GRBs
        \citep{nasa_ws}. The instrument is composed of twelve sodium
        iodide (NaI) and two bismuth germanate (BGO) crystals coupled
        to photo-multiplier tubes. The detectors are mounted on the
        spacecraft in a way that maximizes the overall field-of-view,
        while also guaranteeing that the same event can be seen by
        multiple detectors.
	
	The NaI scintillators detect photons in the energy range of
        $8-1000$ keV while the BGO detectors are optimized for higher
        energies, between $0.2-40$ MeV.  For an in-depth description
        of the instrument performance and the ground calibration of
        the individual detectors, the reader is referred to
        \citet{2009ApJ...702..791M} and \citet{2009ExA....24...47B},
        respectively.

	\begin{figure}[t]
		\hspace{-0.1cm}\includegraphics[width=0.5\textwidth]{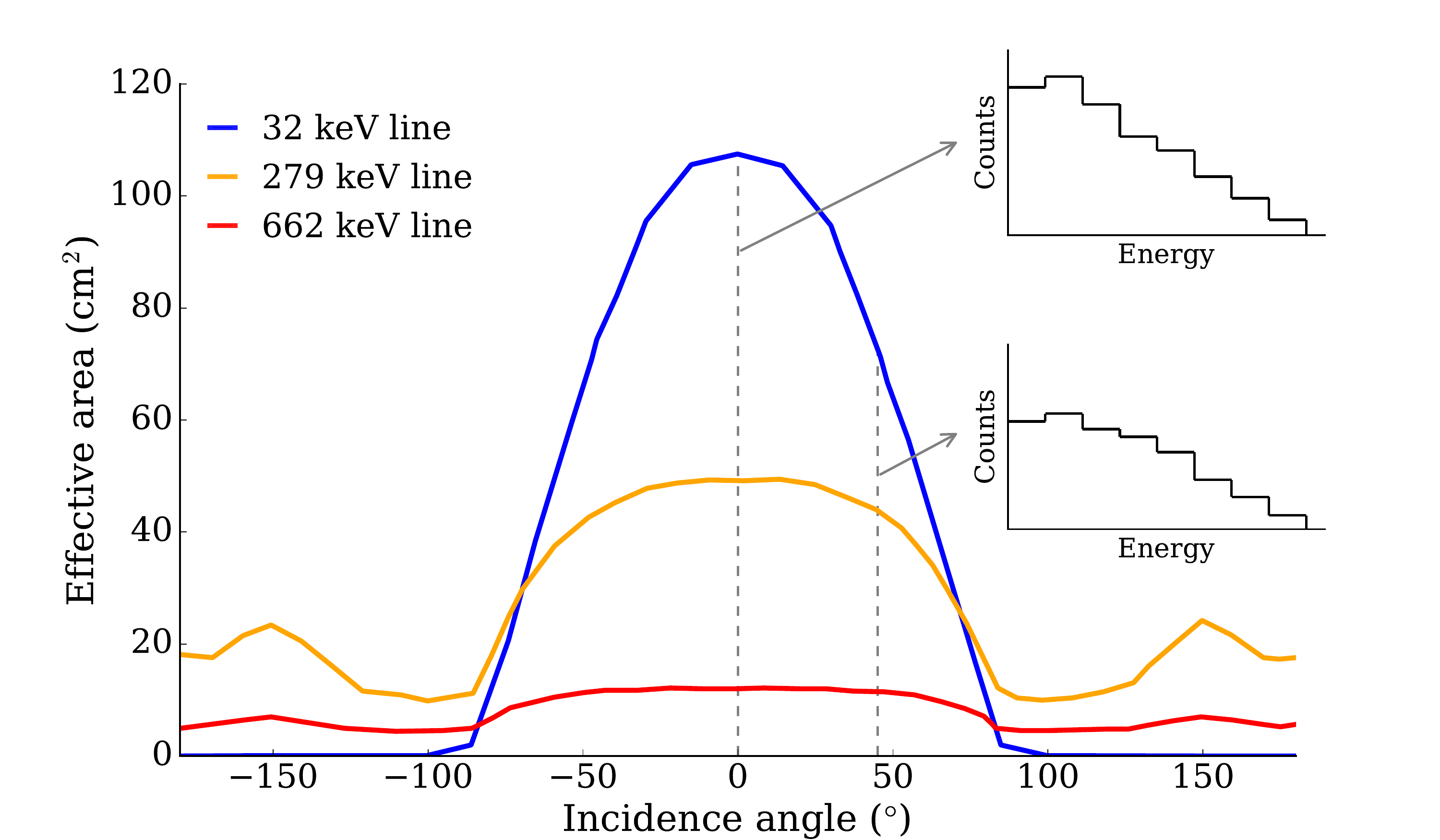}
		\caption{Photopeak effective area for three different energy lines. The response changes drastically with both photon energy and incidence angle. Most of the information about the source location comes from the lower energy channels, where the curves are steeper (i.e. the dependence on the incidence angle is stronger). The figure is a modified version of Fig. 12 from \cite{2009ExA....24...47B}}.
		\label{fig:gbm_photopeak_aeff}
	\end{figure}

	Different data products are regularly generated, but this work
        will focus on the trigdat data in particular, which is the
        only one suitable for rapid localization of GRBs, since no
        other data product is immediately available after the
        detection. For trigdat data the time resolution is finer near
        the trigger time and coarser farther away, ranging from 64 ms
        up to 8.192 s.  The count data are binned in eight energy
        channels for each detector, but only six of them are actually
        usable. For the lowest energy channel, the detector response
        is not very accurate due to photon absorption by passive
        materials, while in the highest channel there are overflow
        issues. This problem arises when the detected photon flux
        exceeds the maximum absorption energy of the scintillators
        and, with its energy only partially measured, this flux
        generates an artificial excess of counts in the last channel.

	Each GBM detector has a response which depends both on the incidence angle and energy of the photons. 
	As shown in Fig. \ref{fig:gbm_photopeak_aeff}, the different shapes of the effective area curves in the plot are due to the change in absorption length for the photons at different energies. In the lower energy channels, the NaI crystal is thick enough to absorb all of the radiation even at normal incidence, when the path of the photon in the crystal is the shortest. As the incidence angle of the photons, $\theta$, grows, the projected area decreases as $\cos \theta$. In the higher energy channels however the crystal is not thick enough to completely absorb the photons due to the rapidly decreasing photoelectric cross-section. As the incidence angle grows the projected area decreases, but at the same time the photons are traveling a longer path in the crystal. These two effects roughly compensate each other and generate the flat effective area curve visible in Fig. \ref{fig:gbm_photopeak_aeff}.

	This particular behavior of the scintillators has the important consequence that lower energy channels are the most sensitive to variations in incidence angle, which means they contribute the most to burst localization (although this is true also because they possess most of the overall count statistics).
	Together with the direct response just described, contributions due to atmospheric and spacecraft scattering of the source's photons are also taken into account in the total response used for event reconstruction \citep{2007AIPC..921..590K}.

	\section{DoL vs. BALROG localizations}
	
	Due to the nature of GBM's response, fully reconstructing a source is a complex task. Spectral shape and position cannot be fitted separately, as they are never independent from each other.
	In principle, there are three possible approaches to source reconstruction:
	\begin{description}
		\vspace{-0.2cm}\item[Case 1] \label{itm:case1} assume that the source is located at some particular fixed coordinates in the sky and then fit for the spectrum.
		\vspace{-0.2cm}\item[Case 2] \label{itm:case2} assume that the source has a particular fixed spectrum, thus fitting only for the position.
		\vspace{-0.2cm}\item[Case 3] \label{itm:case3} make no assumptions and fit for both spectrum and location simultaneously.
	\end{description}

	In \hyperref[itm:case1]{\textbf{case 1}} the source position is assumed to be known due, for example, to simultaneous detection from another mission. If this location is sufficiently precise (i.e. small uncertainties), it is possible to fix the position of the source in the model and just fit the spectrum. While this is a valid approach for spectral fitting, it is of no help when one needs to locate a source with GBM.

	For \hyperref[itm:case2]{\textbf{case 2}} the position is not known, but a particular spectral shape is assumed. This approach has been the standard method to localize bursts with BATSE and GBM for the past 28 years \citep{1999ApJS..122..465P,2016ApJS..223...28N}, with the LOCBURST algorithm for BATSE \citep{1999ApJ...512..362P} and its improved version for GBM, the Daughter of Location (DoL) algorithm \citep{2015ApJS..216...32C}. 
	DoL employs a set of three spectral templates which are generated by making use of the Band function \citep{1993ApJ...413..281B}, and fixing its low energy index ($\alpha$), high energy index ($\beta$) and peak energy ($E_{peak}$) with the values shown in Tab. \ref{tab:band_templates}. The amplitude, which acts as a normalization factor, is the only parameter left free.

	The localization process involves a $\chi^{2}$ minimization for the background subtracted signal \citep{2015ApJS..216...32C}. In practice, this is done by generating a grid of points on the sky, each with the correct response for the specific template spectrum chosen. The $\chi^{2}$ minimization then finds the direction on the sky which most closely matches the measured rates for the detectors.
	It is important to understand that the shape of the spectrum (except the amplitude parameter) is fixed during the whole fitting process, only the positions are being tested. 
	This makes this method unreliable in many cases, whenever the templates 
	do not match the actual burst spectrum.
	Different templates are compared by means of their $\chi^{2}$ and the one with the lowest value is chosen as the best fit model for the GRB.
	Since the three templates are grossly different as compared to the variation
	of the shape of either long- or short-duration GRBs, in general the DoL
	method results in the soft template being primarily providing the best fits 
	for solar flares and Soft Gamma-ray Repeaters, the moderate template for 
	long-duration GRBs, and the hard template for short-duration GRBs.

	\begin{table}[t]
	\centering
	\caption{The Band function templates used in DoL.}
	\label{tab:band_templates}
		\begin{tabular}{c c c c}
			\hline
			\rule{0pt}{3ex} 
			Template & $\alpha$ & $\beta$ & $E_{peak}$ (keV) \\
			\hline
			\rule{0pt}{3ex}
			Soft & -1.9 & -3.7 & 70 \\
			Moderate & -1 & -2.3 & 230 \\
			Hard & 0 & -1.5 & 1000 \\
			\hline
		\end{tabular}
	\end{table}

	Due to the grid discretization, a lower limit of $1^{\circ}$ for the localization error is imposed \citep{2015ApJS..216...32C}. Additionally, the 68\% statistical uncertainty calculated with this method is the average distance to the points in the sky grid with $\Delta \chi^{2} = 2.3$, assuming a circular region for the error.

	Unless there is a more precise location available from another mission, the position generated by DoL is the one provided in the catalogs \citep{2012ApJS..199...18P, 2014ApJS..211...13V, 2016ApJS..223...28N}, and then used for spectral analysis of bursts.
	This represents a circular reasoning: a particular spectral shape is assumed and used to locate the source, but the position obtained is then used to fit the GRB spectrum, which does not generally match the one previously assumed. 
	Whenever the GRB spectrum deviates from the best fit template, this procedure produces a systematic offset proportional to the spectral mismatch, as is evident for bright GRBs also located by {\it Swift} or IPN \citep{2015ApJS..216...32C}.

	A better approach to the analysis is to release the assumption of the source having a particular spectral shape, using the approach defined in \hyperref[itm:case3]{\textbf{case 3}}. This is done by the BALROG (BAyesian Location Reconstruction Of GRBs) code \citep{2018MNRAS.476.1427B}.

	BALROG simultaneously fits both the location and spectrum of the source, generating dynamically the correct detector response without fixing any of the model parameters. The improved computing power now available makes such an approach feasible, while it was impossible during BATSE times: a localization for trigdat data \added{(usually available within 10 minutes)} typically takes about 10 minutes on a multicore workstation.
	In case the selected spectral model does not match the data of the burst sufficiently well, it is always possible to test a different one until a satisfactory result is achieved. Given the present ignorance of the proper physical model of the prompt emission spectrum in GRBs, such iteration can be done only if a better localization of a given GRB is available from a different measurement, e.g. from {\it Swift}.

	\begin{figure*}[t]
          \subfloat{\includegraphics[width=0.51\textwidth, clip=true]{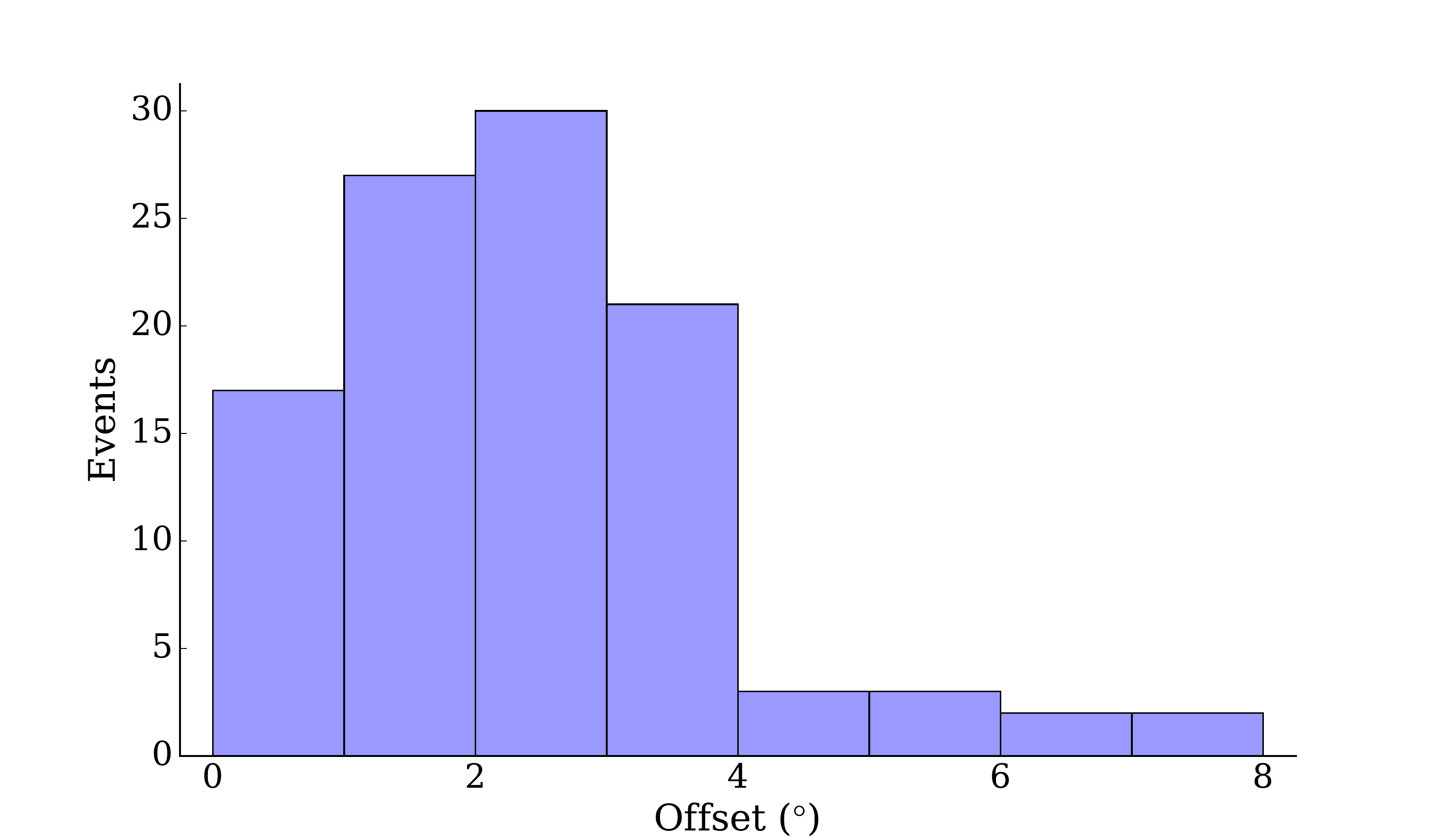} }
          \subfloat{\includegraphics[width=0.51\textwidth, clip=true]{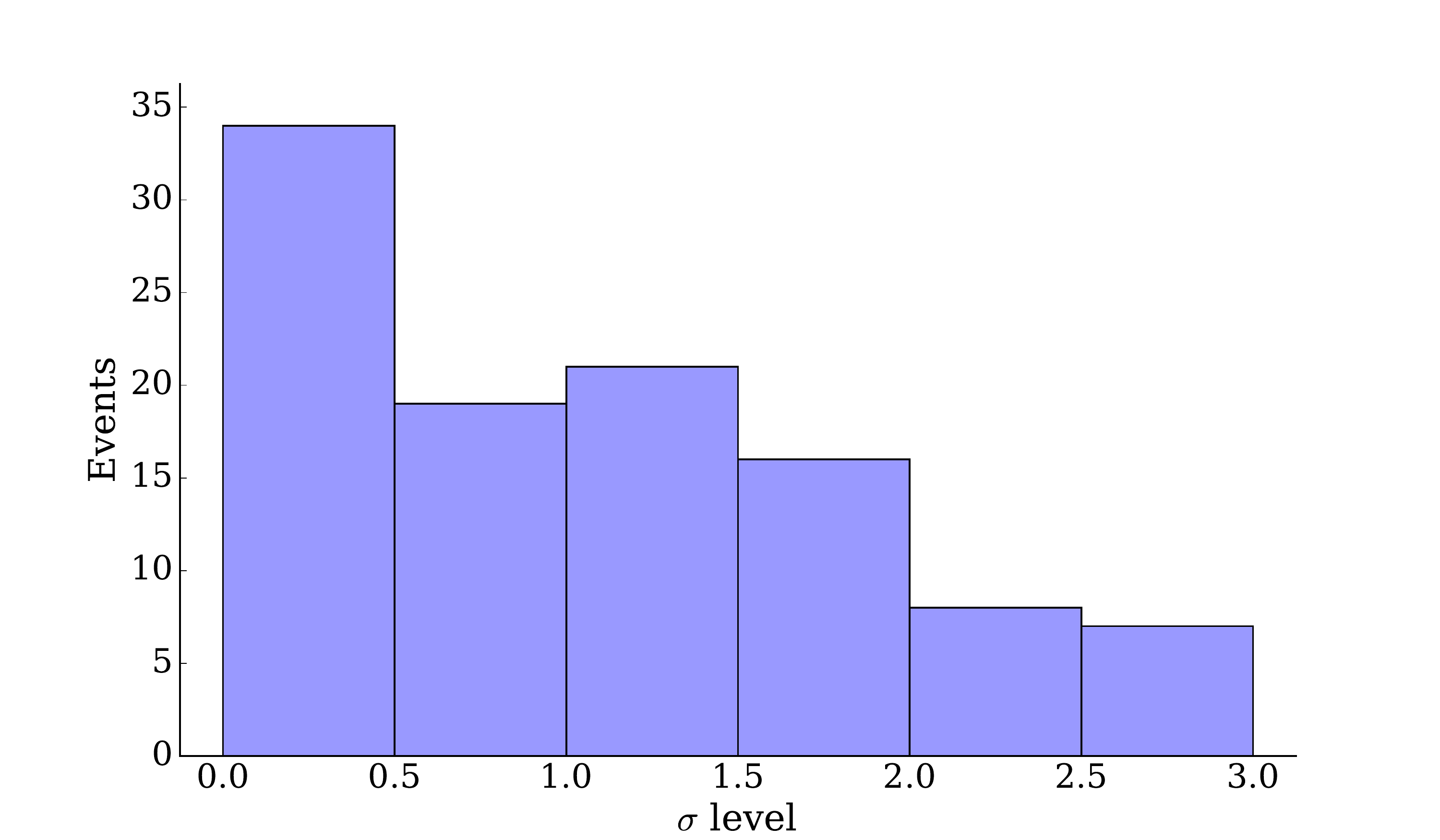} }
		\caption{BALROG offset distribution in units of degrees (left)
                  and sigma level (right).}
		\label{fig:balrog_results}
		\vspace{0.5cm}
	\end{figure*}

	Compared to DoL, BALROG also employs an improved statistical treatment of the background, modeling the counts in each time bin as signal plus background, instead of relying on the statistically incorrect procedure of background subtraction \added{\citep{2018MNRAS.476.1427B}}. \deleted{This allows BALROG to achieve more accurate uncertainties compared to DoL, although this is expected to be more relevant for fainter bursts.
	}

	\section{Analyzed sample and methodology}
	
	Compared to \citet{2018MNRAS.476.1427B}, which introduced BALROG,
	two extensions are implemented
	here: firstly, a larger sample of bright Fermi-GBM GRBs with  {\it Swift} 
	localizations is analyzed, and secondly, a systematic analysis of procedural
	issues is performed. For both, the BALROG scheme is used, and the best fitting 
	spectrum among the following is presented for the final localization: 
	cut-off power law, Band function or a power law. The choice of using these relatively simple spectra is due to the fact that trigdat data is too coarse \added{(6 usable energy channels)} to allow for more complex spectral fitting. With the finer TTE data it is possible to fit more complex spectra, however, since these data products are not readily available, they cannot be used for prompt localization of GRBs. For each fit, 
	all model parameters are left free. The goal is to investigate to which extent 
	the combined fitting of position and spectrum improves the GRB localizations for a larger sample. 

	Only bright bursts have been selected because they have a higher signal-to-noise ratio and can thus be located more precisely. Anticipating
	a much better localization based on \citet{2018MNRAS.476.1427B}, only GRBs 
	are useful for the present study which have statistical localization errors substantially smaller than up to 15\degs\ systematic errors reported in \citet{2015ApJS..216...32C}. \added{Otherwise, no stringent statements on the systematics error can be made}. 
	Only {\it Swift}-localized GRBs are used in this work, since {\it Swift} localizations are much more precise compared to most of the other instruments and can be used as reference to assess BALROG's performance. {\it Swift}'s uncertainties are typically of the order of a few arcminutes for {\it Swift}/BAT \citep{2005SSRv..120..143B} and a few arcseconds for {\it Swift}/XRT \citep{2007SPIE.6686E..07B}. These errors are always much smaller than GBM's and can be neglected when comparing the two positions, thus allowing to reliably assess BALROG's accuracy. 
	\newline
	The aim of this study is twofold: 1) to compare the BALROG localization accuracy with that of DoL and 2) to investigate the offset relative to \textit{Swift} in order to estimate the remaining systematic error. Different GBM detected GRB samples are used for these aims, as defined in the following.

	The sample examined for aim 1) is a subset of what was originally analyzed in \citet{2015ApJS..216...32C}. The smaller sample size used here is motivated by the above described requirement of having small statistical errors.

	A total of 69 GRBs have been analyzed (Tab. \ref{tab:grbs}). Sixty of the bursts were selected by applying a combined cut in the one-second photon peak flux $F_{peak}$ and total fluence $S$, imposing the following:

	\begin{equation}
		F_{peak} > b  - a \cdot S \ ,
		\label{eq:cut}
	\end{equation}

	\noindent
	where $b = 6 \ \rm{cm^{-2} s^{-1}}$ and $a = 0.857 \cdot 10^{5} \ \rm{erg^{-1} s^{-1}}$.
	This particular selection has been made to allow a compromise between having a large enough sample of bright bursts while also avoiding selecting faint or low signal-to-noise GRBs. 
	Additionally, 9 GRBs below the selection threshold have been fitted together with the others.
	\newline
	
	\noindent
	The sample examined for aim 2) consists of all the 69 GRBs from the previous sample, plus another 36 bright GRBs measured after the study done in \cite{2015ApJS..216...32C}. For these bursts there exists a reference \textit{Swift}/XRT position, but an official final GBM localization is not \added{always} public \added{(only the onboard or ground positions are always available)}, so a \added{systematic} comparison between DoL and BALROG positions is not possible for these GRBs. 

	\section{Localization results}
	
	\subsection{BALROG results}
	\label{subsec:balrog_res}

	The result of each BALROG localization is a posterior distribution describing both position and spectral parameters for the chosen model. The final values provided are given by the set of parameters which maximizes the posterior probability distribution. It is, however, possible to obtain much more information than these single point estimates: in our case, it is of particular interest to compute the position distribution for each GRB in the sample. This can be done by marginalizing (i.e. integrating) over the spectral parameter distributions of the model.
	The computed distribution $P(\phi, \theta)$ describes the probability  of the source being located at some particular coordinates $(\phi, \theta)$ on the sky.

	\begin{figure}[t]
        \hspace{-0.5cm}\includegraphics[width=0.56\textwidth]{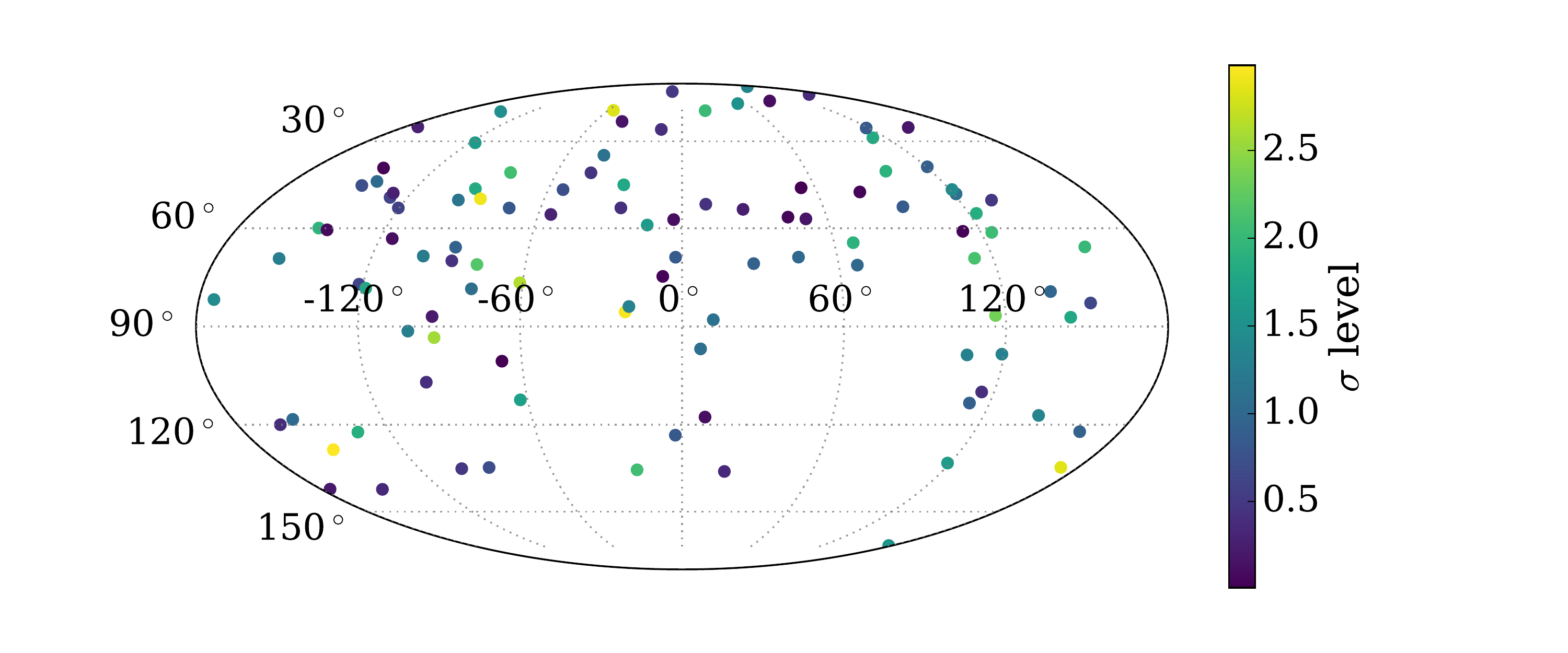}
       \caption{Distribution of the events in spacecraft coordinates, with azimuth angle on the horizontal axis and zenith angle on the vertical axis. In these coordinates, detector b0 lies at $0^{\circ}$, while b1 lies at $180^{\circ}$.}
		\label{fig:sc_skymap}
	\end{figure}

	The results of BALROG localizations are given in 
	Fig. \ref{fig:balrog_results}, with two panels: 
	(i) an offset distribution (i.e. angular separation between fitted and 
	{\it Swift}'s position, left panel) and 
	(ii) an offset distribution in terms of the sigma level (right panel).
	The latter value takes into account the uncertainties associated to each localization and is a measure of statistical accuracy. To understand how this value is computed, consider a source with position probability distribution $P(\phi, \theta)$. The real source position is a single point in the sky (since we neglect {\it Swift}'s position uncertainties) which lies on a contour $C_R$ of equal probability density for $P(\phi, \theta)$. By integrating $P(\phi, \theta)$ over the region inside $C_R$, a probability value $p$ is obtained through the following equation:
	
	\begin{equation}
		p = \int_{C_R}^{} P(\phi, \theta) d\phi d\theta  \ .
	\label{eq:prob}
	\end{equation}

	\noindent	
	This value can be converted to the corresponding number of sigma of a standard normal curve by inverting 

	\begin{equation}
		p = \int_{0}^{\sigma} \frac{2}{\sqrt{2 \pi}} e^{-\frac{x^2}{2} } \ ,
		\label{eq:sigma}
	\end{equation}	

	\noindent	
	where we define $\sigma$ as sigma level.
	More precisely, this curve is half of a normal distribution due to the fact that  offsets are positive definite (i.e. there cannot be a negative angular separation), which implies that the sigma level will also be positive definite.
	For example, a reference location lying on the $68\%$ probability contour region is 1$\sigma$ away, one lying on the 95\% contour is 2$\sigma$ away and so on.
	We decided to adopt the Gaussian formalism for ease of interpretation, it would also have been possible to apply the same procedure with a different distribution.

	In Fig. \ref{fig:sc_skymap}, the distribution of the localized GRBs in spacecraft coordinates is shown. The vast majority of the sources are located in the upper hemisphere, which is to be expected since the GBM detectors are all pointing upwards or horizontally (in spacecraft coordinates). This effect is present due to the fact that a large fraction of the lower hemisphere (in spacecraft coordinates) is occulted by the Earth, although not in a constant way due to the spacecraft rocking.

	\subsection{Comparison between DoL and BALROG locations}
	\label{subsec:dol_comparison}

	In order to investigate the improvement of BALROG localizations, we compare the results here achieved to the ones in \citet{2015ApJS..216...32C}.  
	While computing uncertainties with BALROG locations is trivial once the posterior for the event has been sampled, correctly comparing them with what present in \citet{2015ApJS..216...32C} is not completely straightforward. 
	As already mentioned, the uncertainties provided in the GBM catalogs \citep{2016ApJS..223...28N} are obtained by approximating the $68\%$ probability region (statistical only) to a circle. Since there are no probability contour maps publicly available for the bursts in our sample that date before January 2014, the sigma level for DoL is computed as the offset of the achieved localization to \textit{Swift}'s divided by the statistical uncertainty provided.

\begin{figure*}[th]
    	\subfloat{\includegraphics[width=0.51\textwidth]{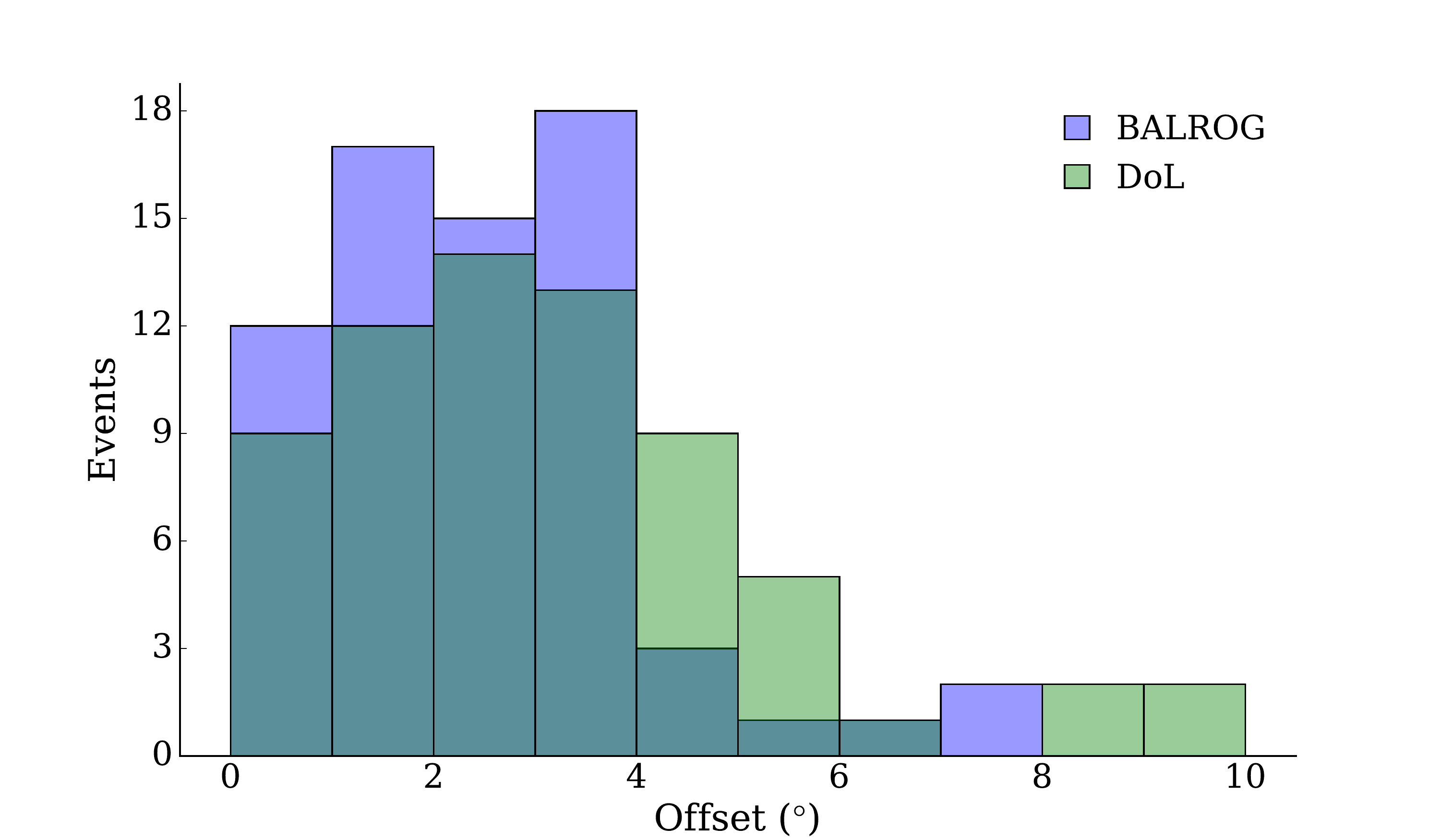} }
    	\subfloat{\includegraphics[width=0.51\textwidth]{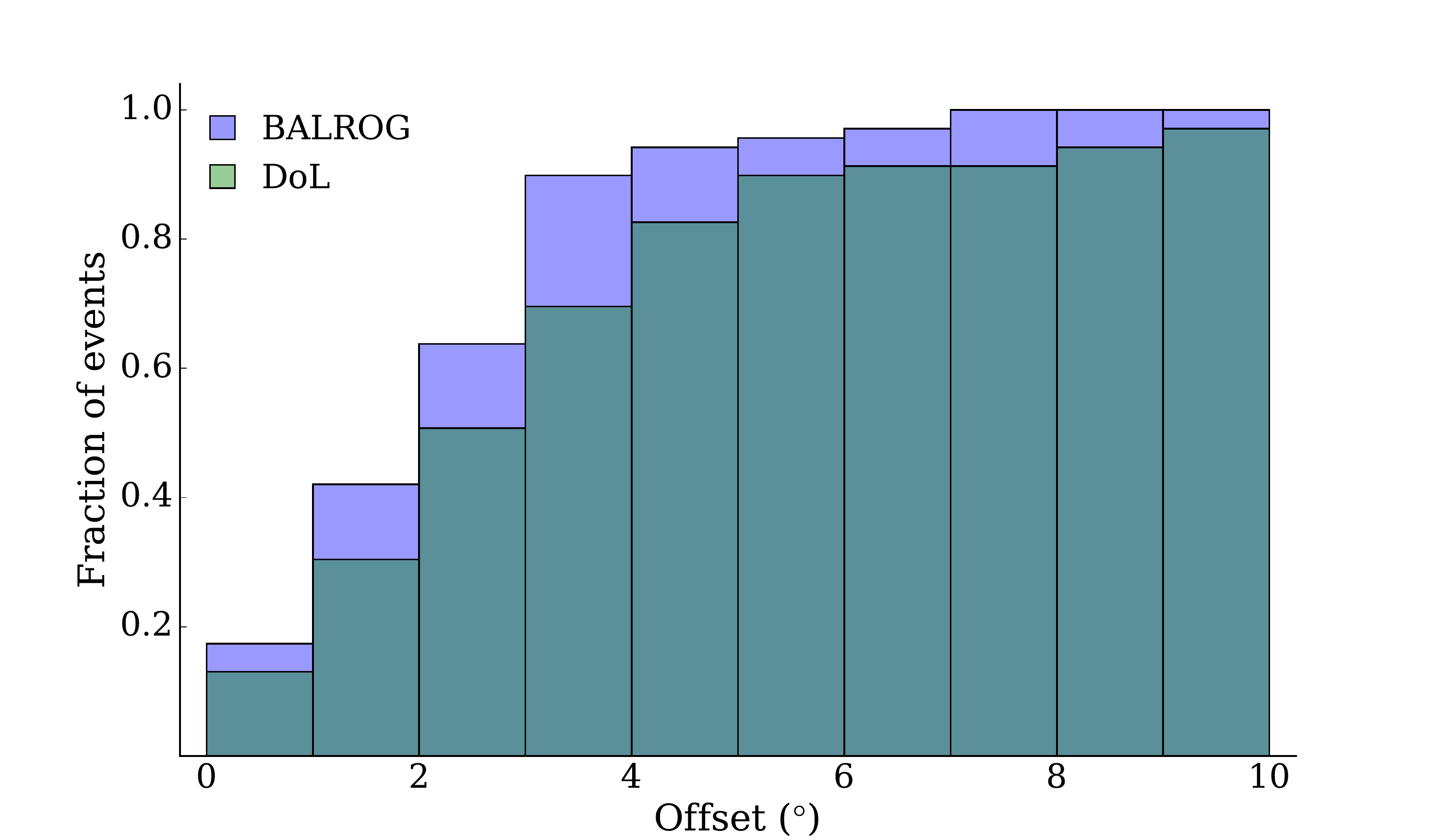} } \\\vspace{-0.3cm}
		\subfloat{\includegraphics[width=0.51\textwidth]{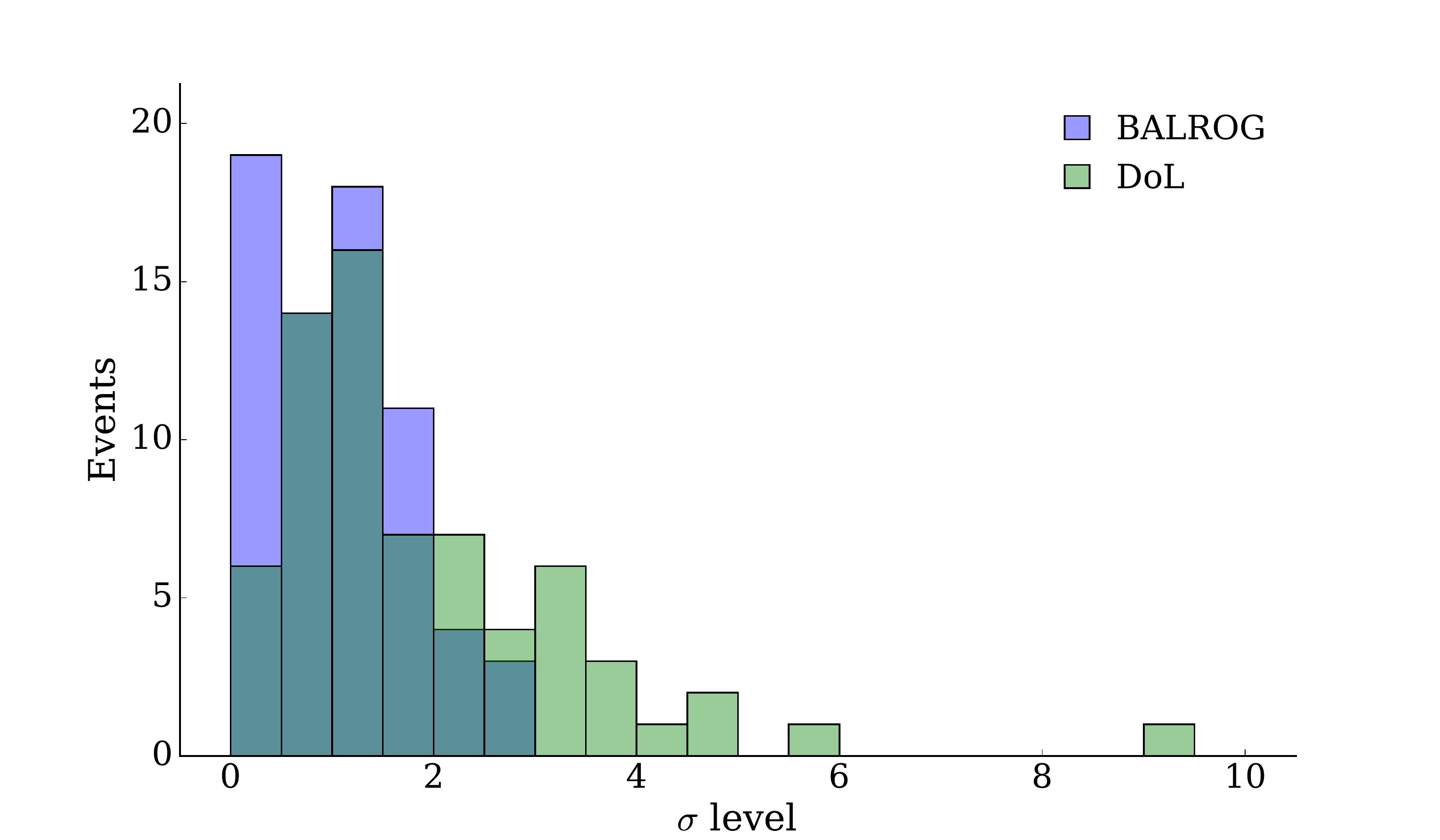} }
        \subfloat{\includegraphics[width=0.51\textwidth]{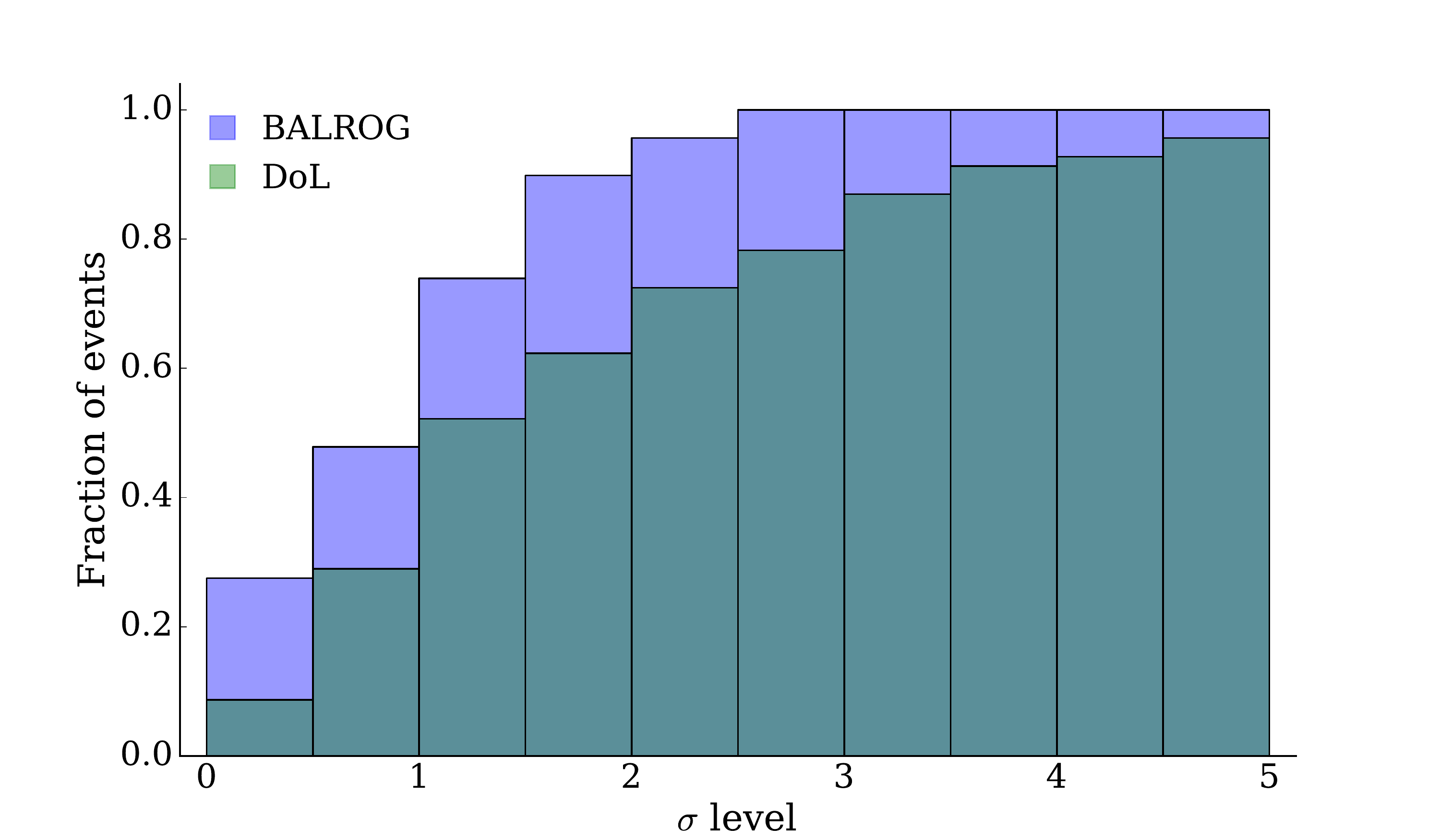} }
        \vspace{5mm}
		\caption{BALROG and DoL localization performance comparison, again
        separated for offsets in degrees (top) and sigma level (bottom). The left column shows the histogram distribution, while the right shows the cumulative distribution. Two DoL localizations with offset $\sim 16^{\circ}$ and $\sim 19^{\circ}$ are not shown in the offset plots.}
		\label{fig:balrog_dol_comparison}
	\end{figure*}

	The performance of each burst localization can be evaluated through two simples quantities. One is the offset, which quantifies how large the angular distance between fit and reference location is. The second quantity is the sigma level, which quantifies how off the fit position is by also taking into account uncertainties. Using this as a method of comparison, BALROG substantially improves the accuracy of the localizations (Fig. \ref{fig:balrog_dol_comparison}).
	In particular, localizations at very large sigma level are completely removed with BALROG, while with DoL many of such outliers were still present. Bursts with very large offset ($>$10\degs) are also removed with BALROG, while with DoL there are two of such occurrences, with angular separations of $\sim$16\degs\ and 
	$\sim$19\degs\ (not shown in the plots). 
	
\begin{figure}[th]
		\hspace{-0.4cm}\includegraphics[width=0.51\textwidth]{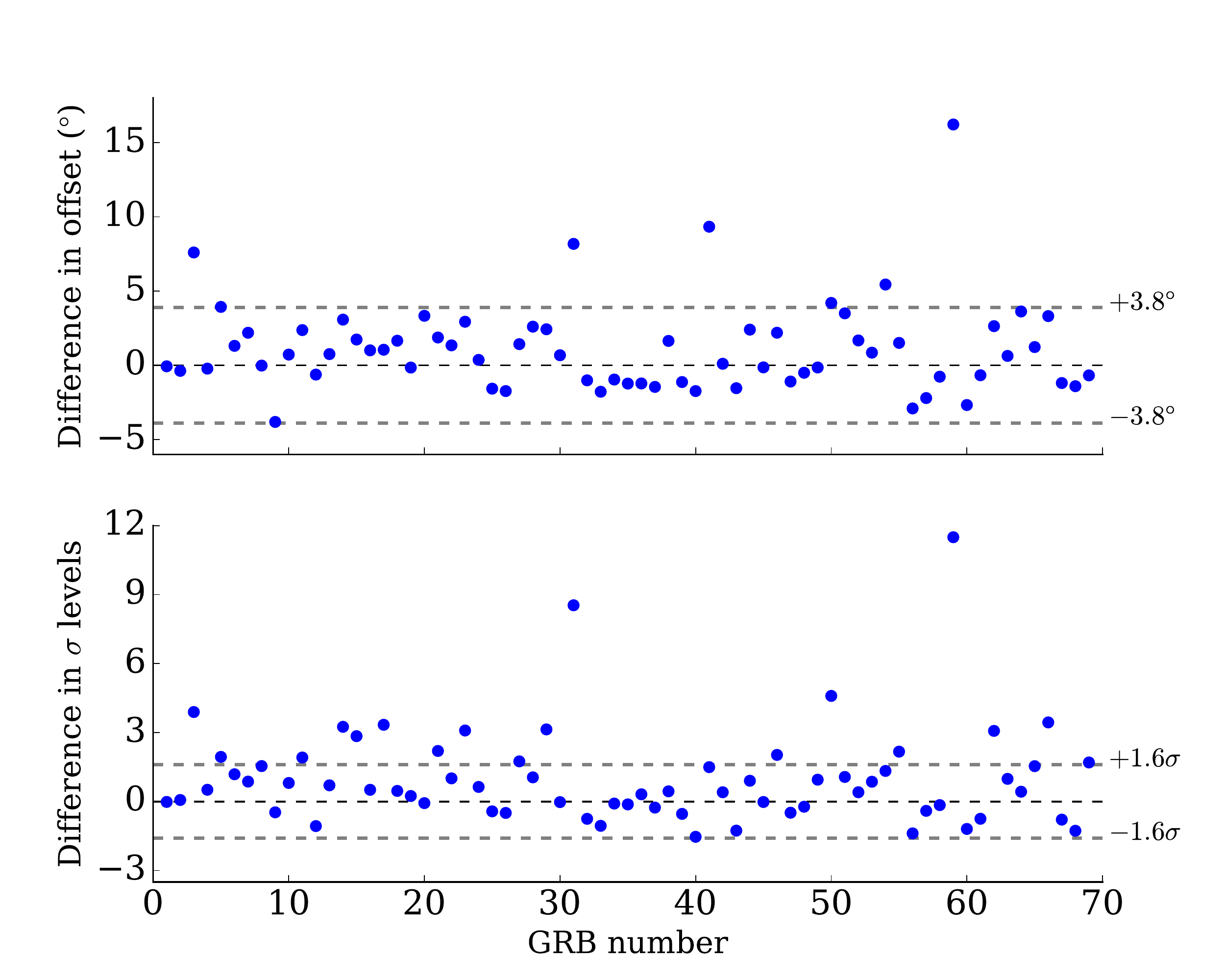}
		\caption{Difference of the offsets produced by the two codes DoL and BALROG, 
                 in units of degrees (top) and sigma level (bottom). In terms of accuracy, 18/69 GRBs (26\%) are badly localized only by DoL, while BALROG never produces positions which have offsets beyond the 3$\sigma$ credible region.\\}
		\label{fig:stat_scatter}
	\end{figure}

	In Fig. \ref{fig:stat_scatter}, the difference in performance between DoL and BALROG is shown. A running number is assigned to each burst and the difference in terms of sigma levels for the two codes is visible. The bulk of the sample (51/69 = 74\%) shows similar performance for the two algorithms, with some scattering due to statistical noise, i.e. for half of those the BALROG positions are 1--2\degs\ larger, for the other half 1--2\degs\ smaller than with DoL. However, a 26\% fraction of the sample (18/69) is poorly localized only by DoL, whereas BALROG is able to reliably locate such bursts and never results in offsets larger than the 3$\sigma$ statistical error. It is this fraction of inaccurate DoL locations which force the GBM team to convolve the statistical error with a systematic one \citep{2015ApJS..216...32C}.

	\subsection{Localizations with Band function templates}

\begin{figure*}[ht]
		\subfloat{ \includegraphics[width=0.51\textwidth]{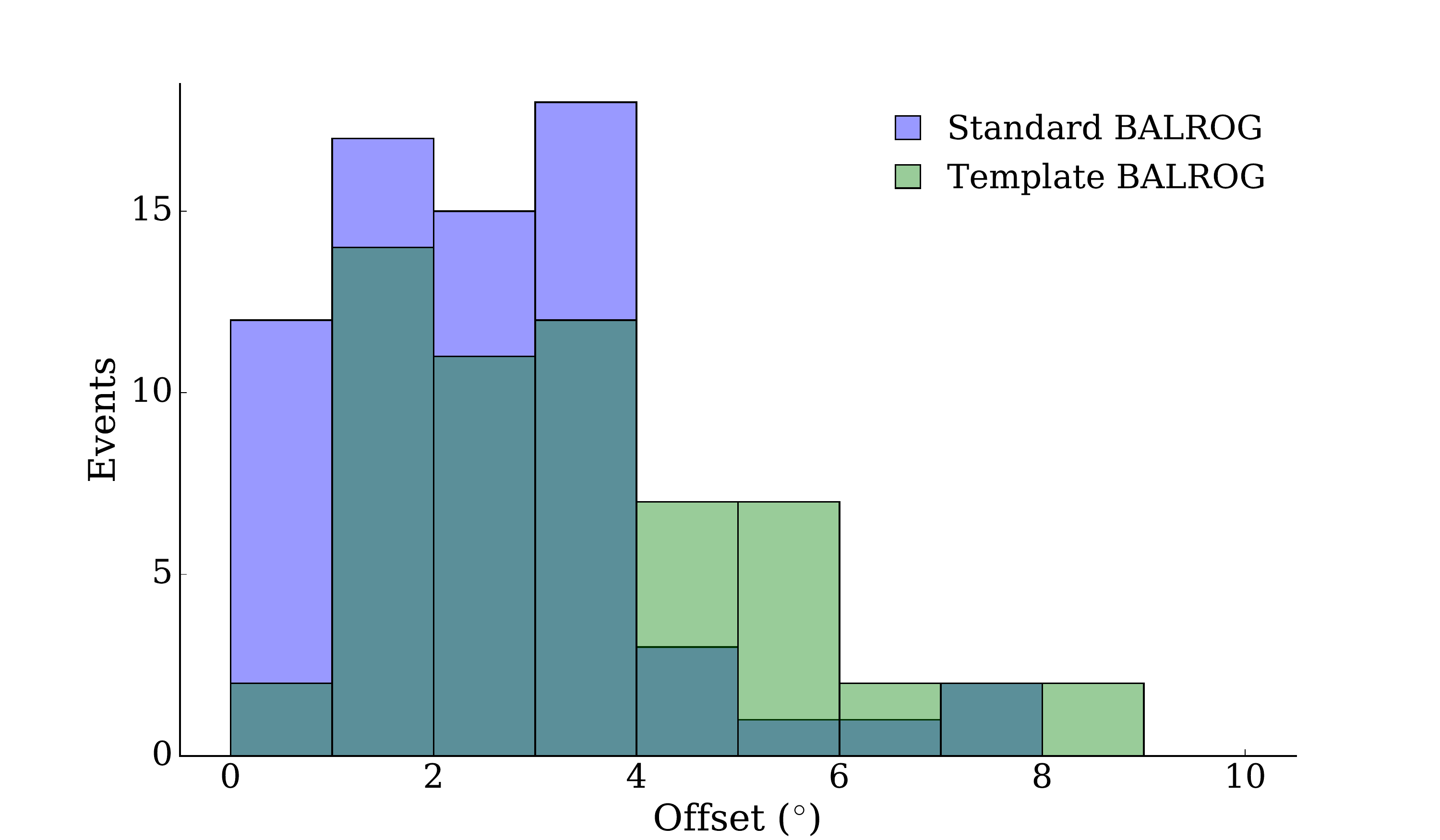} }
		\subfloat{\includegraphics[width=0.51\textwidth]{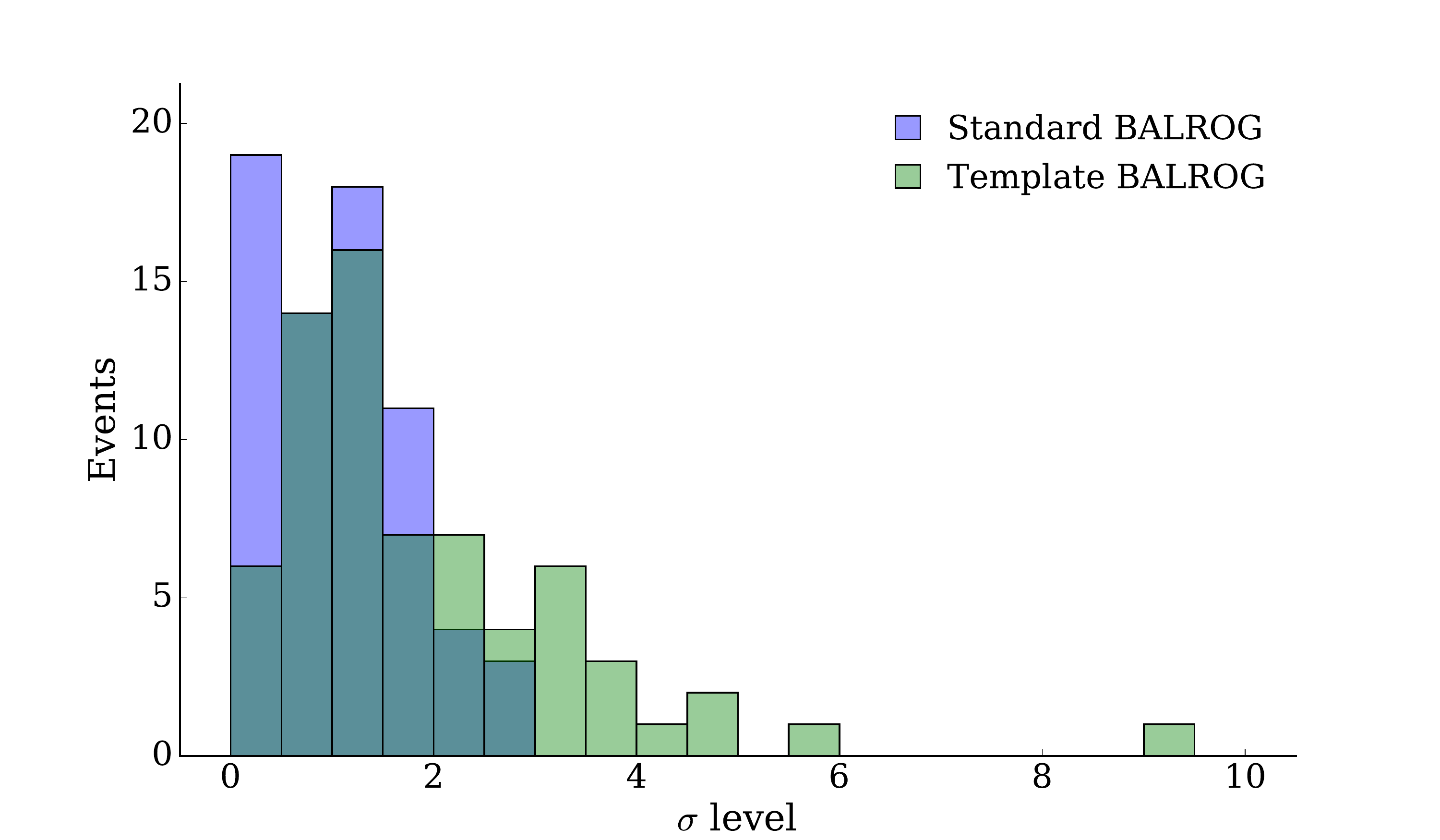} }\\
		\vspace{-2mm}
		\caption{BALROG localization comparison with free and template spectral models. On the left panel, the offset distributions are plotted in degrees, while on the right panel in the sigma levels.}
		\label{fig:balrog_band_comparison}
	\end{figure*}

	It is of interest to quantify how much difference there is between using a free spectral model (i.e. not a particular model with fixed parameters) over a Band template with BALROG. To make the comparison as fair as possible, the same Band templates used by \citet{2015ApJS..216...32C} have been adopted and fitted again to all the 69 bursts of the first sample. Once again both offset and sigma level distributions are compared (Fig. \ref{fig:balrog_band_comparison}).

	The plots clearly show that allowing a free spectral model increases both accuracy and precision compared to using template Band functions. This proves our earlier conjecture, because without the use of a free spectral model the algorithm employed cannot generate a truly dynamical response. This implies that often locations obtained through the use of (a few) fixed templates are unreliable due to the poor match between template spectrum and the actual source spectrum.

	\subsection{Accuracy of BALROG's error contours}
	\label{subsec:accuracy}

\begin{figure}[t]
		\hspace{-0.1cm}\includegraphics[width=0.48\textwidth]{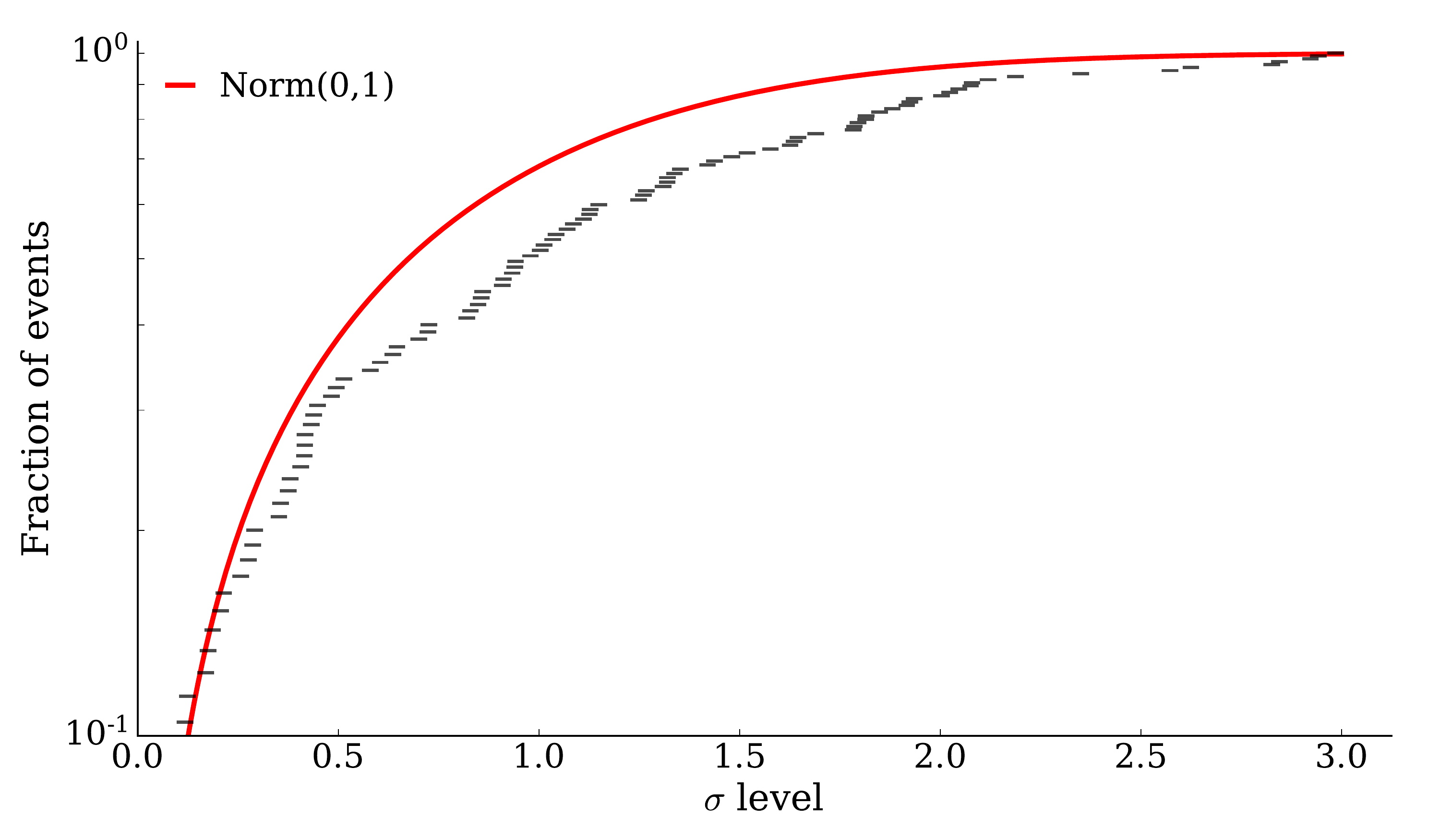}
		\caption{Comparison between achieved (black lines) and the expected (red line) \added{BALROG} CDF values (expressed in log scale).}
		\label{fig:balrog_norm_cdf_sigma}
	\end{figure}

	It is important to verify whether BALROG's error contours are accurate or not, that is to check if the credible regions include the true position of the source as often as they should.
	If this happens, it means that there are no significant systematics left and that the inaccuracies in the DoL localizations are primarily not due to issues in the response (e.g. incomplete modeling of the spacecraft photon scattering, inaccurate model for atmospheric scattering etc.), but are instead only caused by a wrong methodology in the reconstruction of the source. 
	If the error contours are accurate, the cumulative distribution function (CDF) should reach 68\% at 
	1$\sigma$, 95\% at 2$\sigma$, and so on.
	In Fig. \ref{fig:balrog_norm_cdf_sigma} the expected and achieved CDFs are compared and a discrepancy between the two is visible.

	In the following sections, we conduct an in-depth study of where this discrepancy may arise from.

	\section{Search for systematics}	

	BALROG already achieves a significant improvement in terms of localization quality. In the following, the effect of a number of (sometimes implicit) assumptions on the resulting error distribution is investigated to see if further improvements are possible now that any inaccuracy due to wrong fitting methodology is removed. 
	In this section, we use sample 2 with 105 GRBs.

	\subsection{Earth and Sun separation}
	
	Gamma-ray emission from Earth and Sun can potentially affect the quality of the localizations due to enhancing the background.

\begin{figure}[t]
		\hspace{-0.1cm} \includegraphics[width=0.48\textwidth]{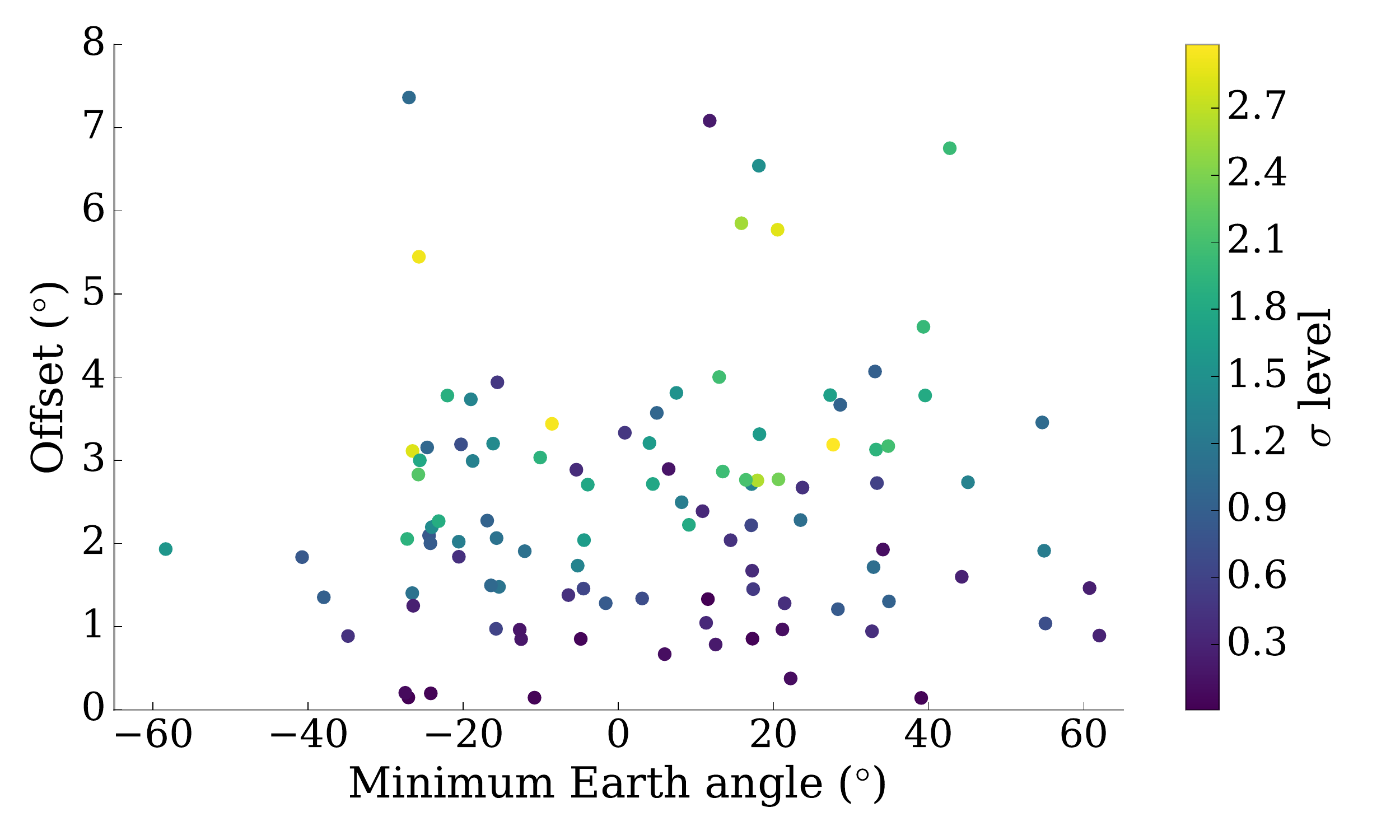}
		\caption{Angular separation between the detector boresight and
		the Earth's limb vs. localization quality, showing no systematic trend. 
		Negative angles denote detectors pointing towards Earth, i.e. below
		the Earth's limb as seen from the spacecraft.}
		\label{fig:earth_angle}
	\end{figure}

	The Earth is a bright gamma-ray source in GBM's energy window \citep{2008ApJ...689..666A} and could in principle decrease the precision and/or accuracy of the localizations. The Earth constantly occults a large portion of GBM's field-of-view and it is rather common to have one or more detectors with signal from the burst which are also facing the Earth.
    Fig. \ref{fig:earth_angle} shows the minimum angle between the detectors used for the fit and the Earth's limb, verifying that if the background is properly fitted (and the atmospheric response is accurate) there is no  dependence of the localization accuracy and Earth albedo 
	emission.
	
	\begin{figure}[h]
		\hspace{-0.1cm} \includegraphics[width=0.48\textwidth]{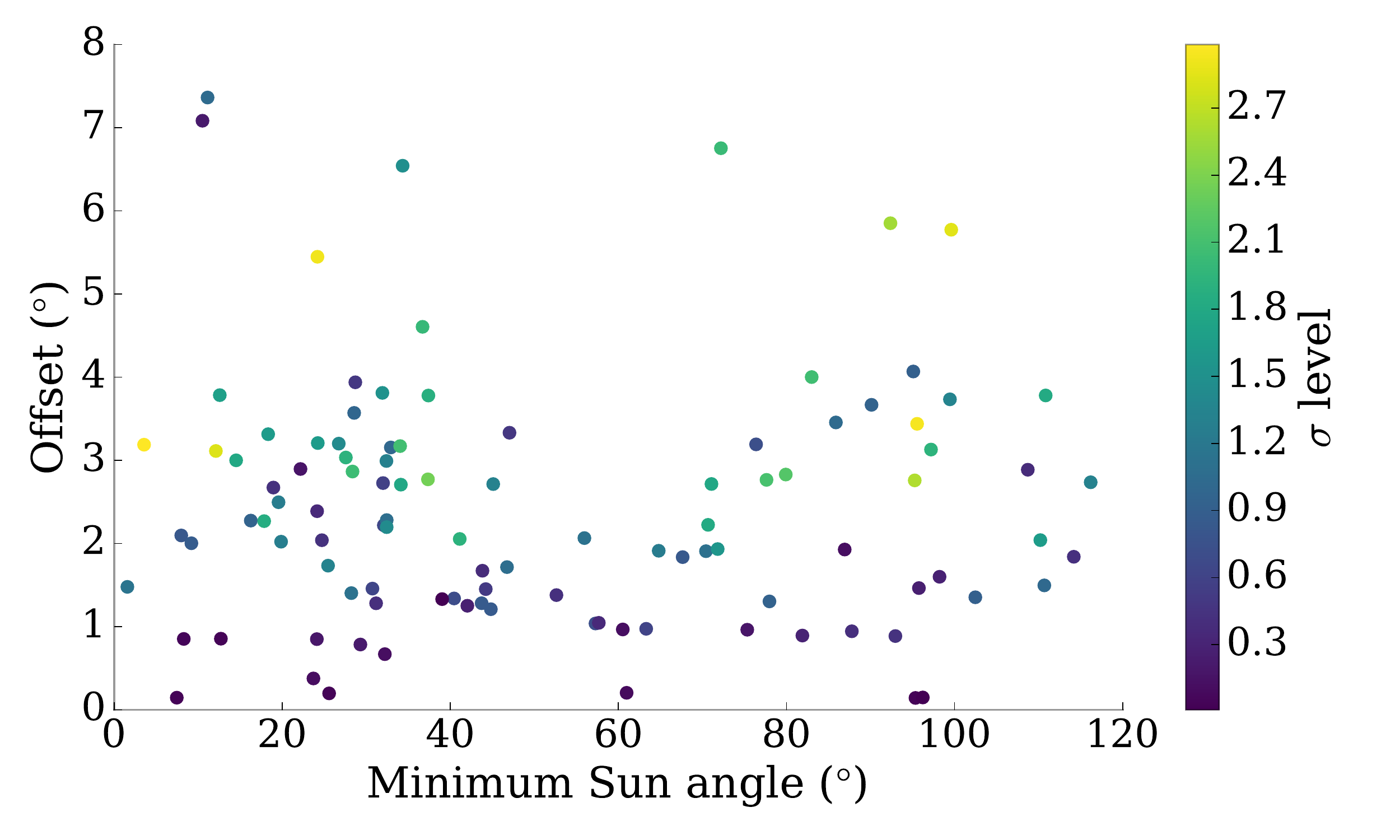}
		\caption{Angular separation between the detector boresight and
			the Sun vs. localization quality, showing again no systematic trend.}
		\label{fig:sun_angle}
	\end{figure}

	Similarly, the Sun does not affect the localization quality, as shown
	in Fig. \ref{fig:sun_angle}, where the minimum separation angle between the detectors used for the fit and the Sun is plotted.
	The position of the Sun on the sky is roughly tracked by detector n5.

	\subsection{Detector pairs}
	\label{subsec:det_pairs}
	
\begin{figure*}[th]
        \hspace{-0.4cm}\includegraphics[width=0.37\textwidth]{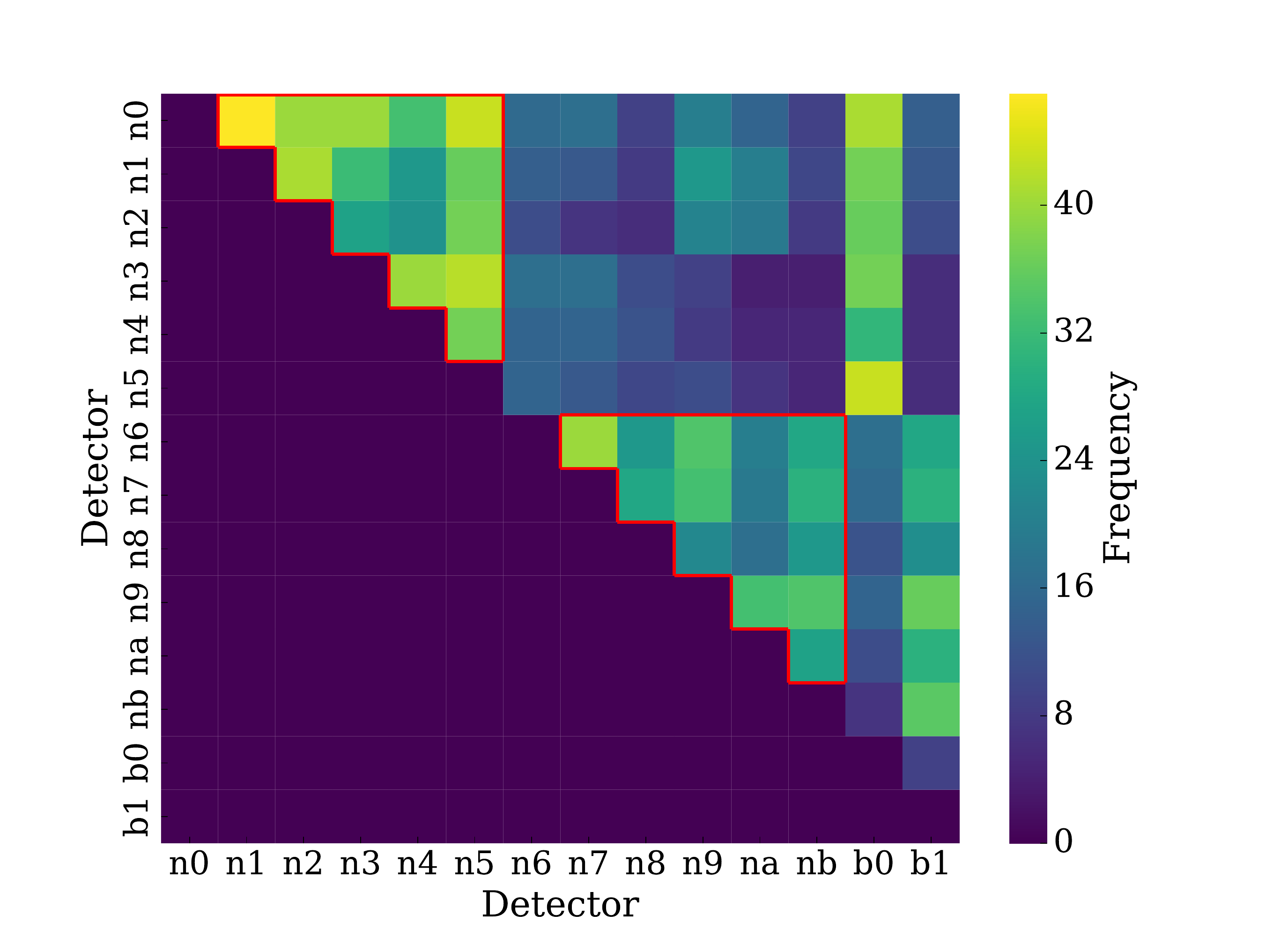}
        \hspace{-0.6cm}\includegraphics[width=0.37\textwidth]{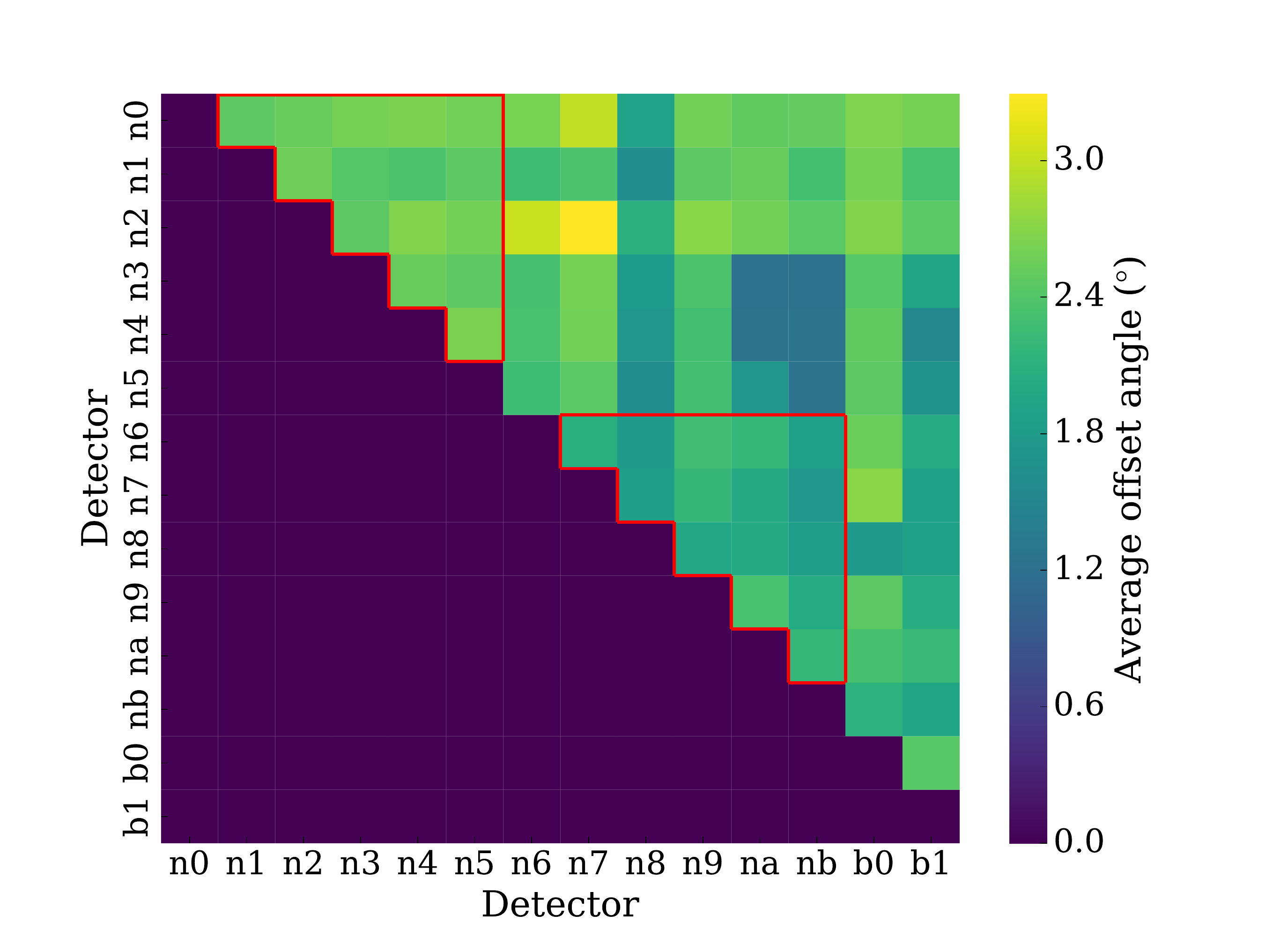}
		\hspace{-0.6cm}\includegraphics[width=0.37\textwidth]{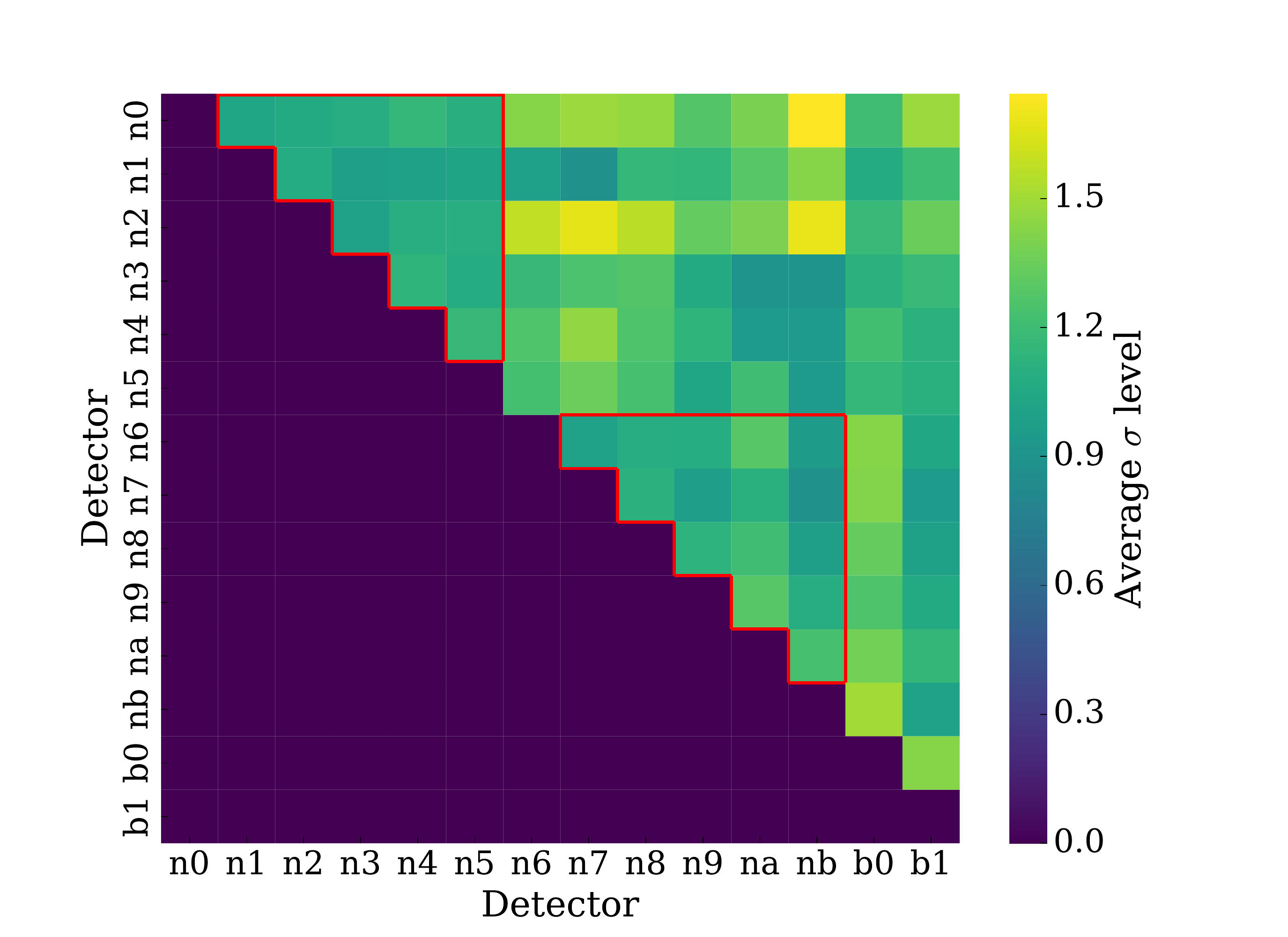}
		\caption{Count and performance of detector pairings: 
        	Left: Frequency with which each pair of detectors is used for fitting. Due to the geometry of the spacecraft, some detector combinations are used more frequently than others (red contour line).
        	The average offset for each detector pair is shown in degrees (Middle) and in sigma level (Right).}
		\label{fig:frequency_pairs_grid}
	\end{figure*}

	While the performance of the single detectors is well understood from ground calibrations \citep{2009ExA....24...47B}, it is nonetheless interesting to verify if particular detector pairs, i.e. just specific combinations of two detectors, perform worse or better than others in terms of localization accuracy. For each of these detector pairs the number of times they are used for fitting is counted.
	From a simple geometric consideration it is clear that there are two detector groups which work together frequently, namely from n0 to n5 plus b0 (b0 side of the spacecraft) and from n6 to nb plus b1 (b1 side), see Fig. \ref{fig:frequency_pairs_grid}.
  	The reason for this is purely assembly geometry, as the two different subsets correspond to the opposite sides of the spacecraft where the detectors are mounted, and detectors on the same side of the spacecraft are much more likely to detect the same burst.

	To see if any pair is under-performing, one can check if any particular combination is systematically less accurate and/or precise. This is done by computing the average offset achieved by each pair, as shown in the middle and right panels of Fig. \ref{fig:frequency_pairs_grid}.
	As one can see, the b0 side of the spacecraft is performing worse than the other. This difference can be seen in both panels, which implies that those localizations are both less precise and less accurate.
	This asymmetry in the performance of GBM will be investigated further in the next subsections.

	\subsection{Dependence on spacecraft coordinates and detector sets}
	\label{subsec:det_sets}
	
	The design task of GBM on the Fermi-satellite was to observe the part of the sky which is
	not covered by the LAT. Since for Fermi's orbit the Earth covers nearly
	half the sky towards nadir, the GBM detectors are oriented such that 
	zenith (LAT boresight) and directions below the Earth horizon are
	underweighted.
	In this subsection, the dependence of the localization quality on the spacecraft coordinates will be looked at in some more detail.

	\begin{figure}[bh]
		\hspace{-0.01\textwidth} \includegraphics[width=0.48\textwidth]{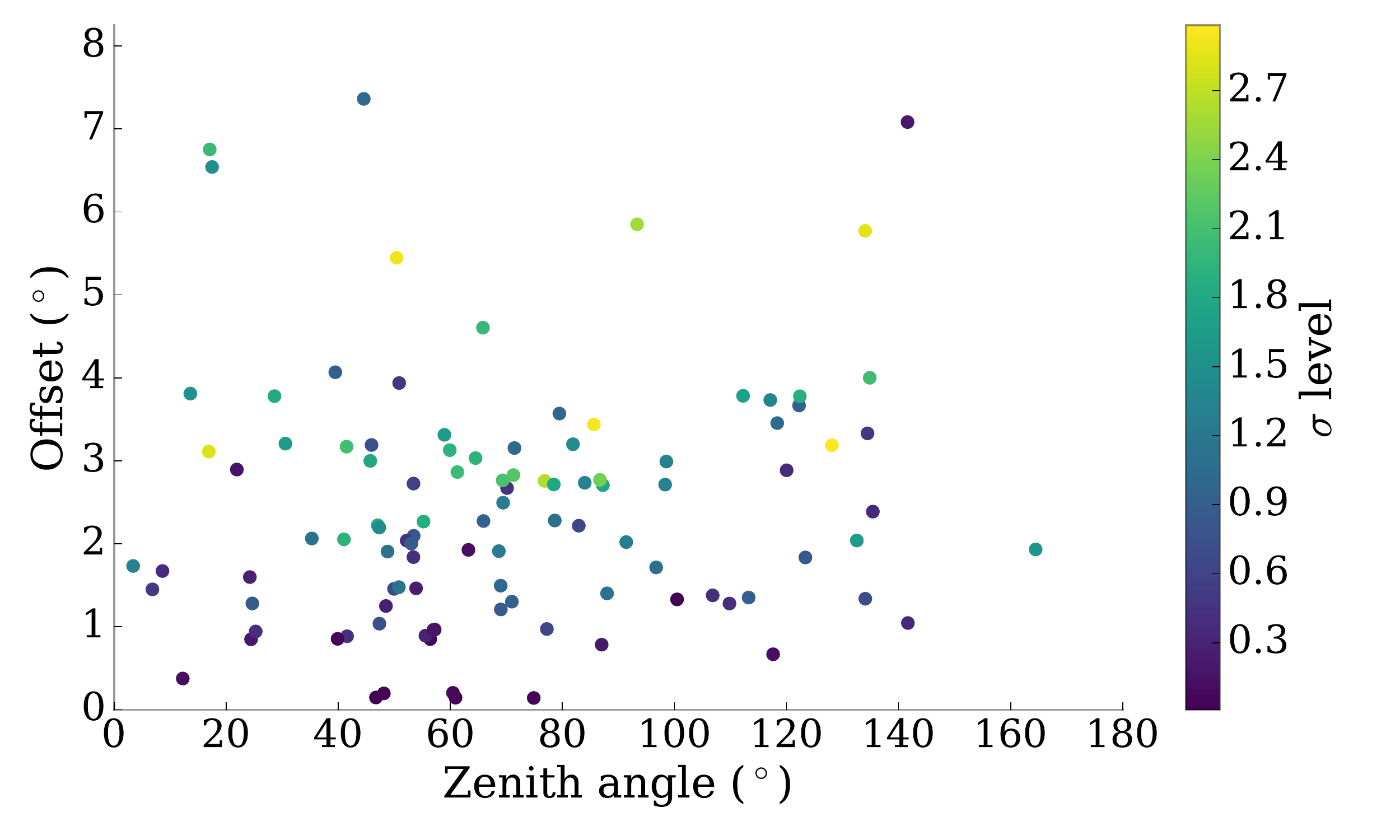}
        \caption{Offset variations depending on the spacecraft zenith coordinate for each event. $0^{\circ}$ denotes the spacecraft pointing direction.}
		\label{fig:zenith_distr}
	\end{figure}

\begin{figure}[h]
		\hspace{-1cm} \includegraphics[width=0.56\textwidth]{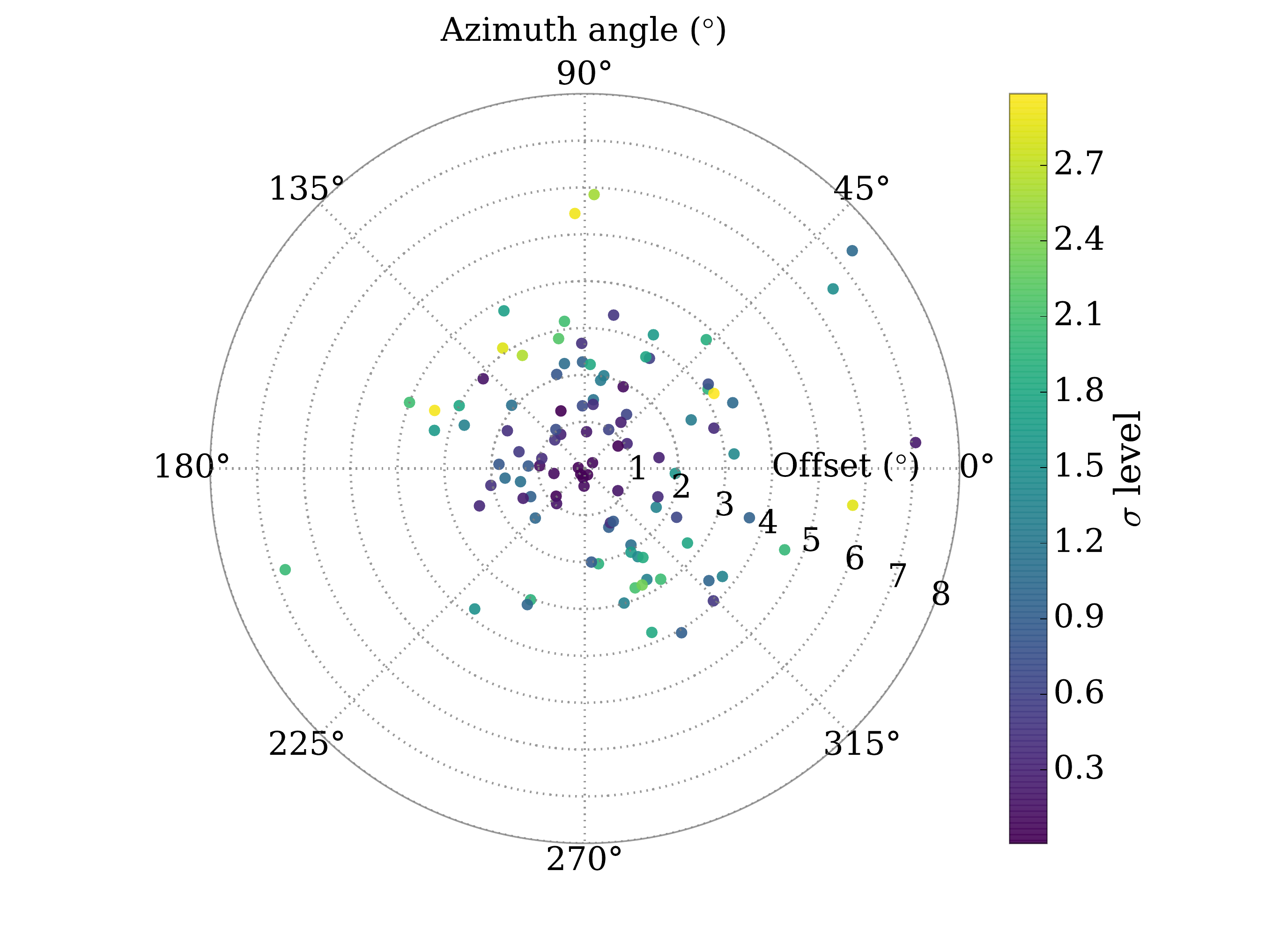}
		\caption{Offset and sigma level variations depending on the spacecraft azimuth coordinate for the event. Azimuth angles of 0\degs and 180\degs correspond respectively to detector b0 and b1.}
		\label{fig:polar_azimuth_delta}
	\end{figure}

	In Fig. \ref{fig:zenith_distr}, the distribution of the GRBs of our sample with respect to the spacecraft zenith angle is shown.
	Localizations in the upper hemisphere of GBM are overall better, which is to be expected given that the detector array geometry is optimized to locate bursts coming from those directions.
	In Fig. \ref{fig:polar_azimuth_delta}, the dependence of the localization quality on the spacecraft azimuth angle is shown, and once again it is easily noticeable that one side (b1) of spacecraft is both more accurate and precise than the other one (b0).
                                                                                                                                                                                                                                                                                                                                                                                              
\begin{figure*}[t]
		\hspace{-1.3cm} \includegraphics[width=0.56\textwidth]{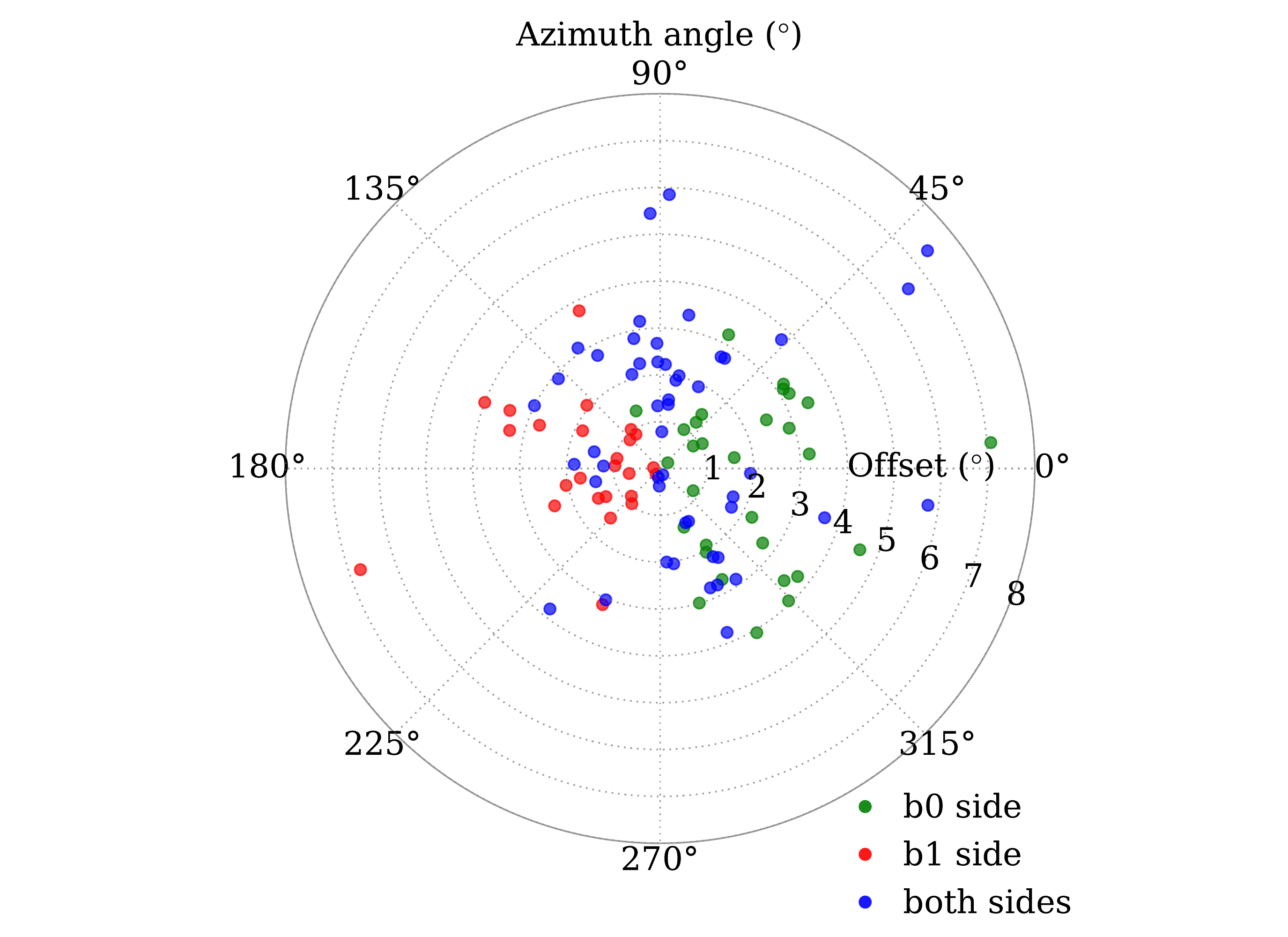}
        \hspace{-1.3cm} \includegraphics[width=0.56\textwidth]{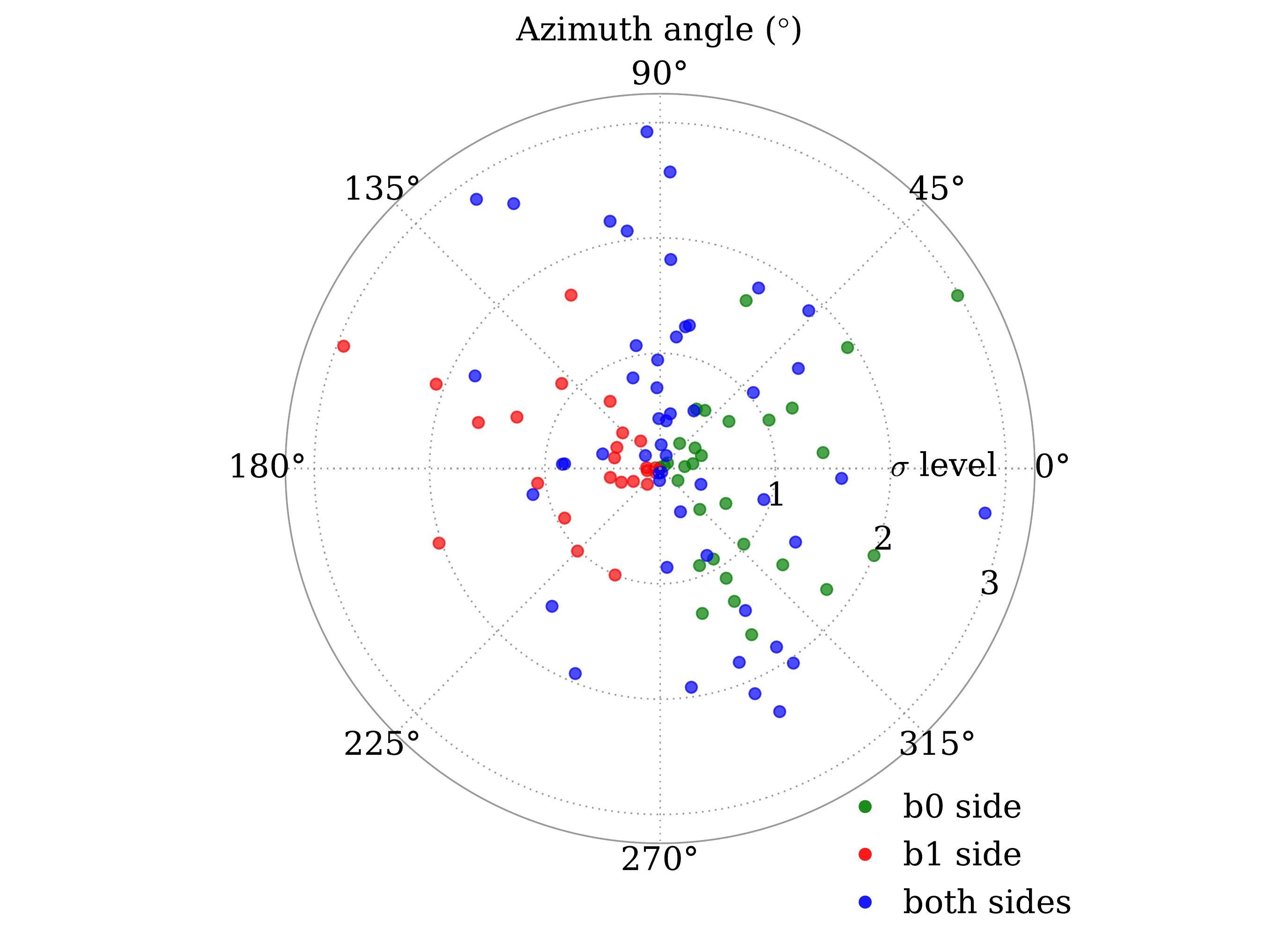}
		\caption{Offset variations in degrees (left) and sigma level (right) depending on the spacecraft azimuth coordinate and detector set for each event. Solar panels and radiator plates are situated at 90\degs/270\degs.}
		\label{fig:polar_azimuth_offset_sets}
	\end{figure*}

	This difference in performance can be made even more evident by splitting the full sample in three distinct subsets, dividing the events in bursts detected only by detectors on one specific side of the spacecraft and bursts detected by both sides. There are thus three subclasses of events: b0 side (30 events), b1 side (25 events) and both sides (50 events), as shown in the two panels of
	Fig. \ref{fig:polar_azimuth_offset_sets}.

	\begin{figure*}
		\subfloat{\includegraphics[width=0.48\textwidth]{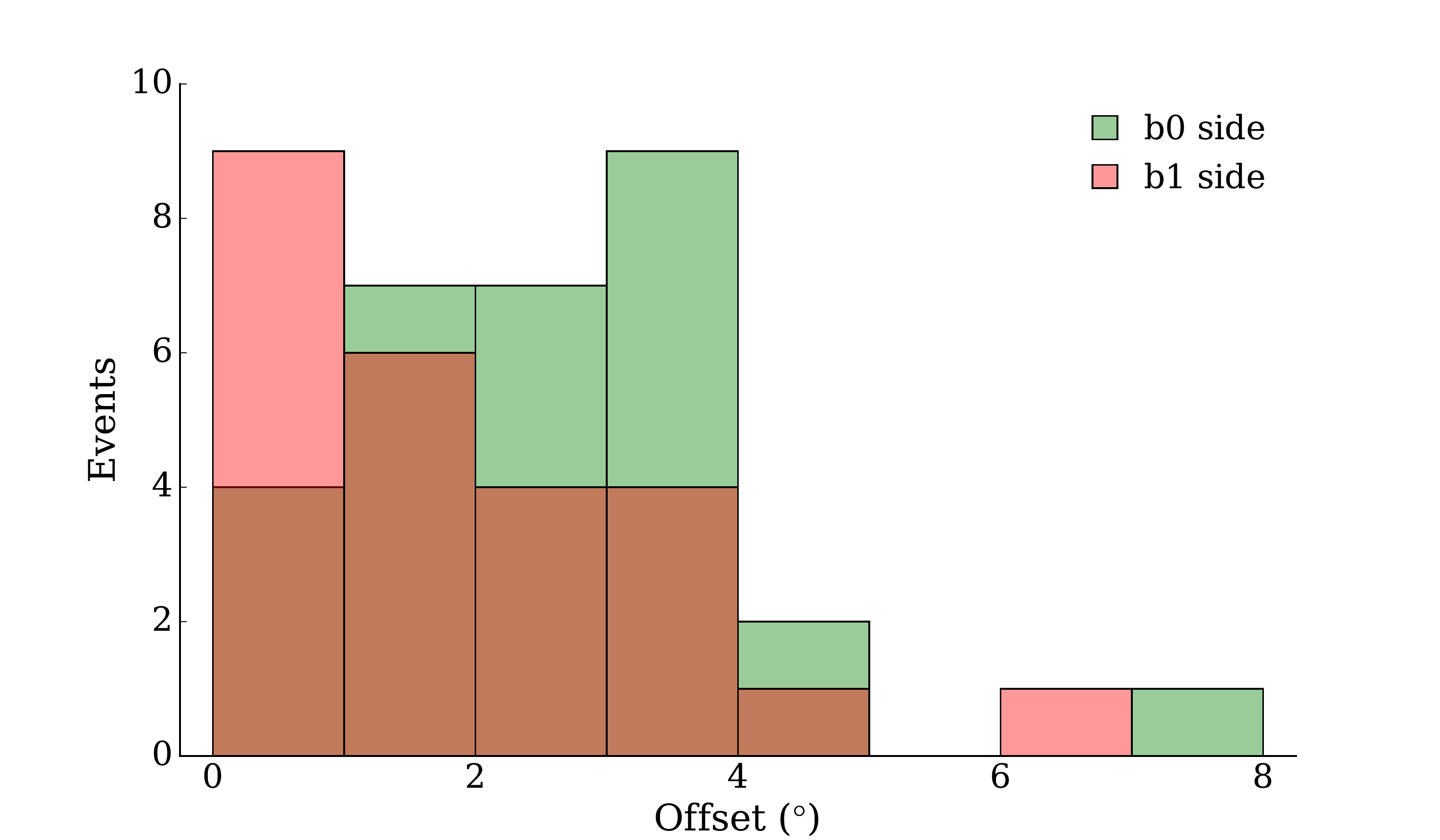} }
		\subfloat{\includegraphics[width=0.48\textwidth]{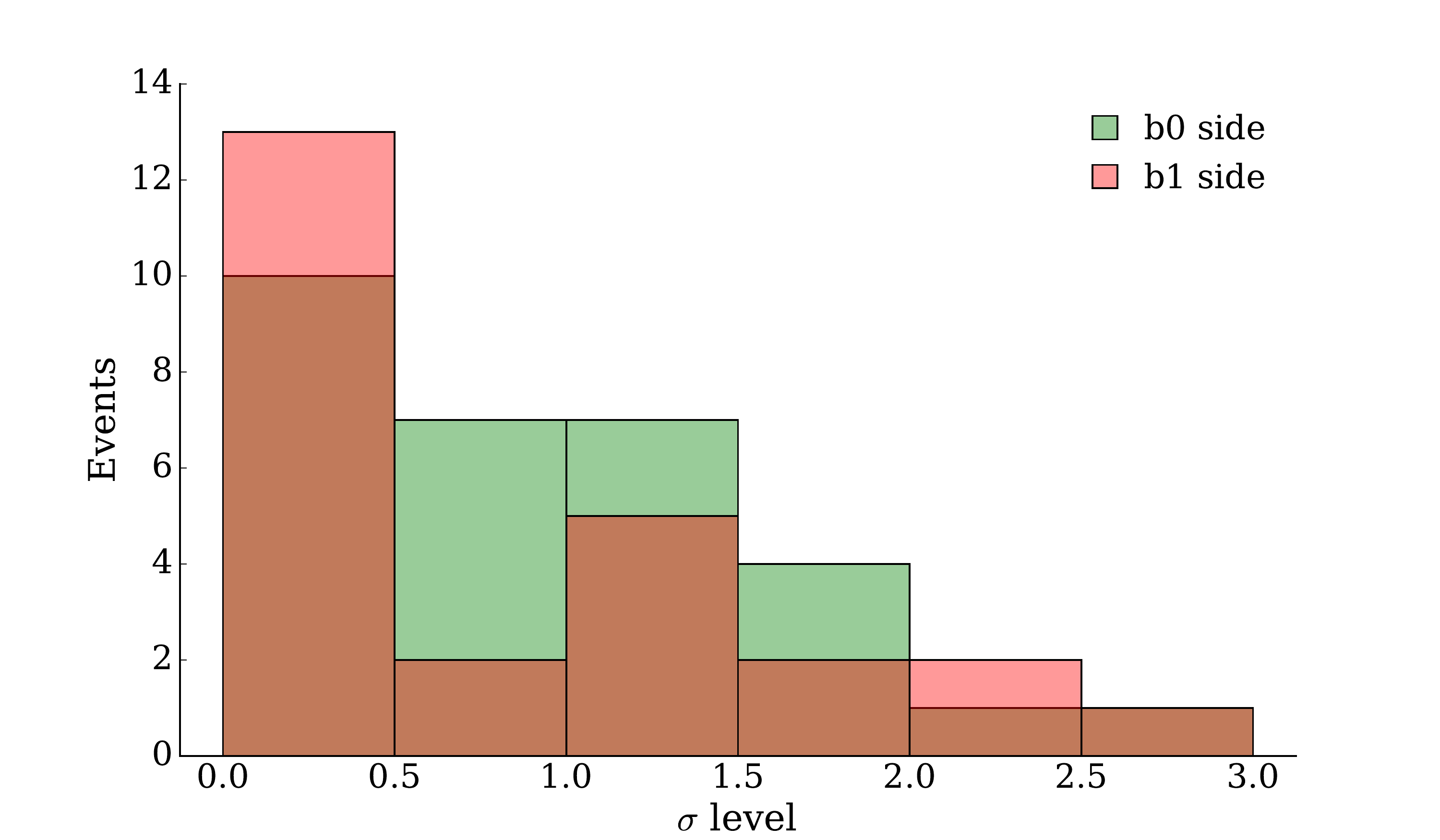} }
		\caption{Offset (left panel) and sigma level (right panel) distribution comparison for the two opposite sides of the spacecraft.\\}
		\label{fig:comparison_sets}
	\end{figure*}

	Offsets on the b0 side of the spacecraft are noticeably worse, as visible also from the histogram version in Fig. \ref{fig:comparison_sets}. Interestingly, localizations made using detectors from both sides of the spacecraft are also worse than the ones made by either the b0 or the b1 side. This may hint to the presence of some unmodeled or inaccurate scattering response for e.g. the solar panels, which are in fact located at spacecraft azimuth angles 90\degs\ and 270\degs.

\begin{figure}[ht]
		\hspace{-0.01\textwidth} \includegraphics[width=0.48\textwidth]{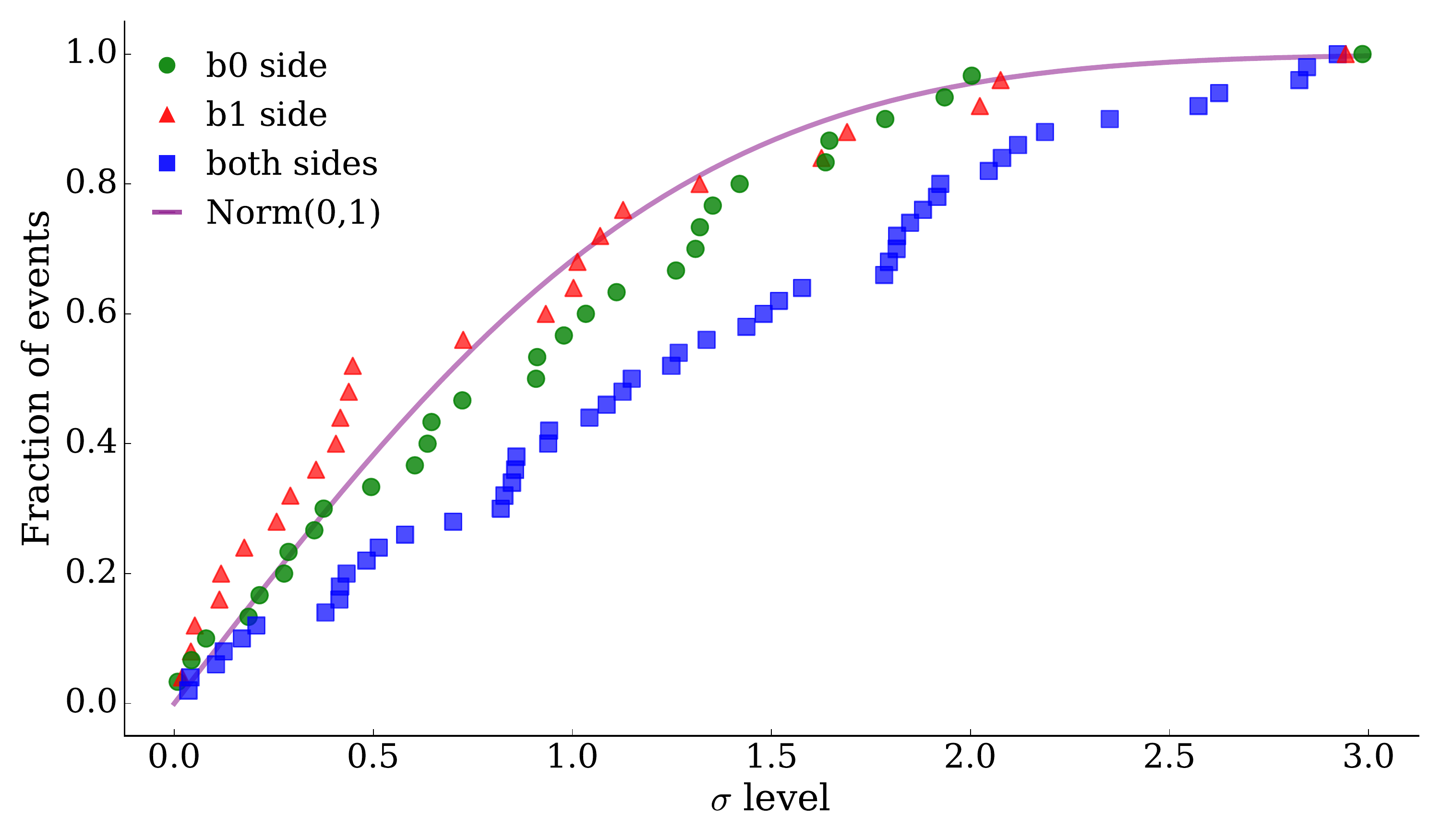}
		\caption{Sigma level CDF comparison for the three different subclasses.}
		\label{fig:cdf_sets}
	\end{figure}

	In Fig. \ref{fig:cdf_sets}, the CDFs of the three different subsets are compared and they present a significantly different behavior. In the large numbers limit, all the subsets should by construction follow a standard Gaussian distribution CDF, if there are no systematics. Particularly noticeable is the fact that locations fitted with detectors from both sides seem to be the least accurate ones.

	\subsection{Offset distribution}

\begin{figure}[bh]
		\hspace{-0.01\textwidth} \includegraphics[width=0.54\textwidth]{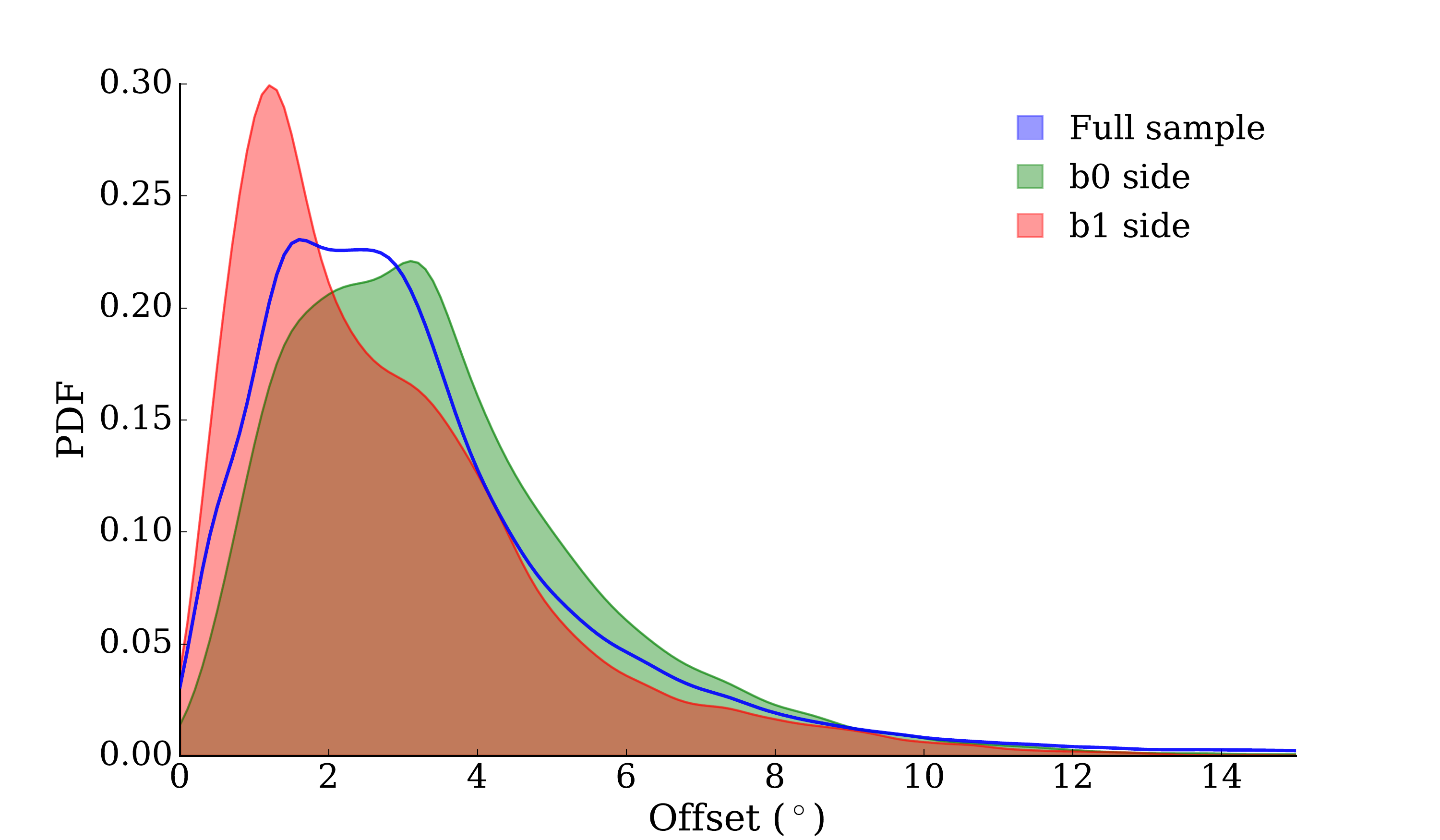}
		\caption{Offset distributions for the two opposite sides of the spacecraft (green and red). The blue line shows the offset distribution for the total sample, including the 50 GRBs seen from both sides.}
		\label{fig:offset_distr}
	\end{figure}

	For each fitted event, the posterior distribution allows us to analyze the two-dimensional position distribution on the sky after marginalizing over the spectral parameters. From this, one can easily compute a distribution of the offset for that source from a certain position in the sky (e.g. a {\it Swift} location), which defines the probability of having a certain offset angle for a particular source.
	This procedure can be repeated for each single GRB in our sample. In principle, each of these offset distributions is specific only to a particular source spectrum and geometrical configuration of the spacecraft (i.e. where the source is located in spacecraft coordinates, which detectors are occulted and so on). 
	However, each of these distributions can be considered as a sample coming from a global offset distribution, which represents the overall localization performance of the spacecraft for bright bursts (Fig. \ref{fig:offset_distr}).

	
	In the same way as before, the distribution can be split in three distinct subsamples depending on the side of the spacecraft used for the localization (b0, b1 or both of them). 
	From Fig. \ref{fig:offset_distr}, one can once again see that the b0 side of the spacecraft is achieving a lower performance in terms of localization, consistent with subsections \ref{subsec:det_pairs} and \ref{subsec:det_sets}. Additionally, there appears to be a secondary peak or shoulder in both of the distributions, which is much more prominent for the b0 side curve. This hints at some kind of systematic effect which is always present, but to a different extent on the two sides.

	\subsection{Statistical noise and accuracy}
	\label{subsec:mc_sim}
	
	Finally, it is worth to verify whether the datasets from the sample are compatible with the hypothesis of no systematic error, that is to check if statistical noise alone is enough to explain the observed behavior.
	In the case of no systematics, the CDFs  shown in Figs. \ref{fig:balrog_norm_cdf_sigma} and \ref{fig:cdf_sets} are expected to  follow the CDF of half a standard normal, in the limit of large numbers. 
	The sample analyzed is, however, rather small and statistical fluctuations are not negligible when comparing the actual data to the expected distribution.

\begin{figure}[ht]
		\hspace{-0.01\textwidth} \includegraphics[width=0.50\textwidth]{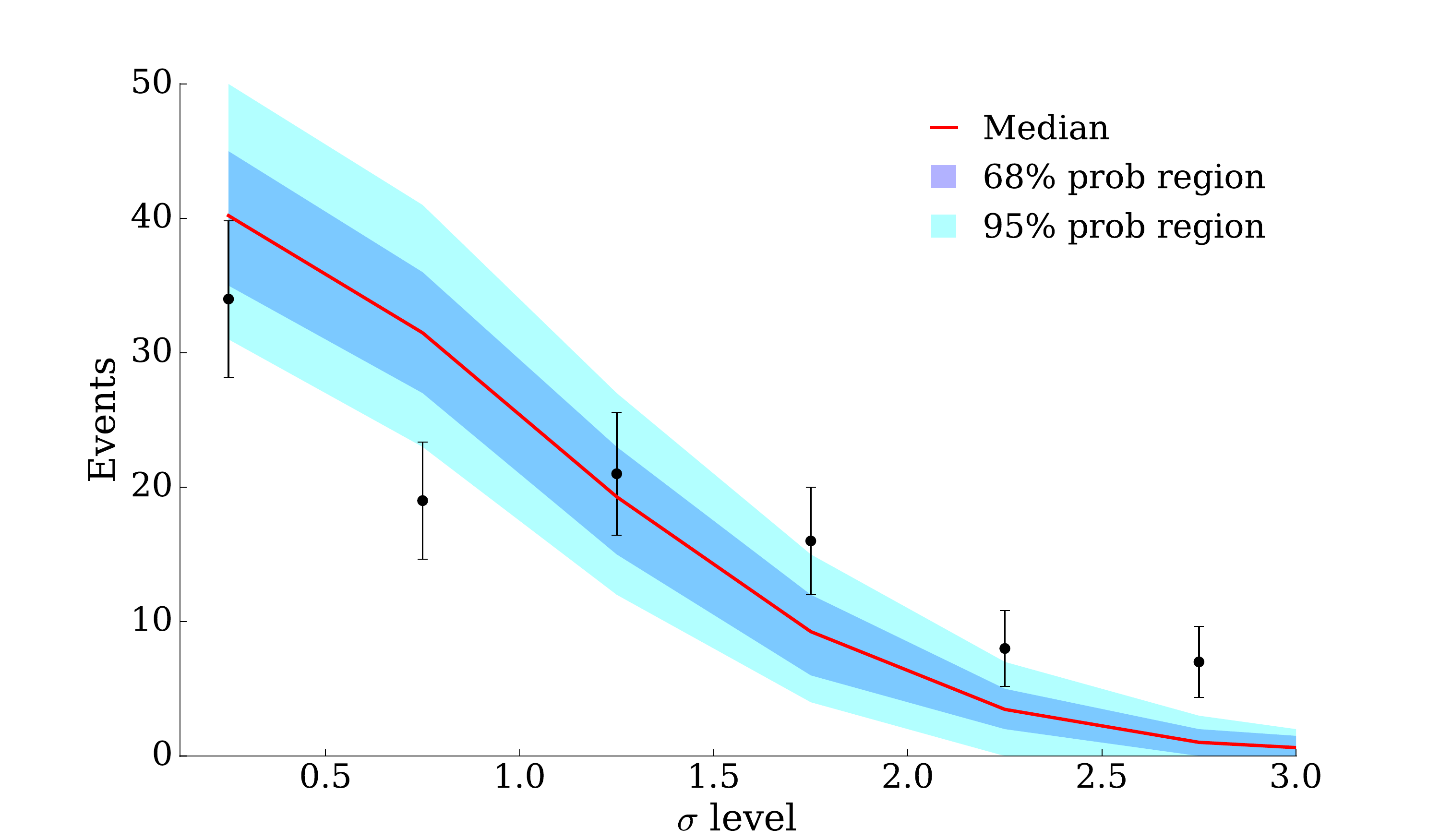}
		\caption{Comparison between the percentile regions for the simulated datasets and all of the BALROG data points (105 GRBs).}
		\label{fig:mc_datasets}
	\end{figure}

\begin{figure}[ht]
		\hspace{-0.01\textwidth} \includegraphics[width=0.50\textwidth]{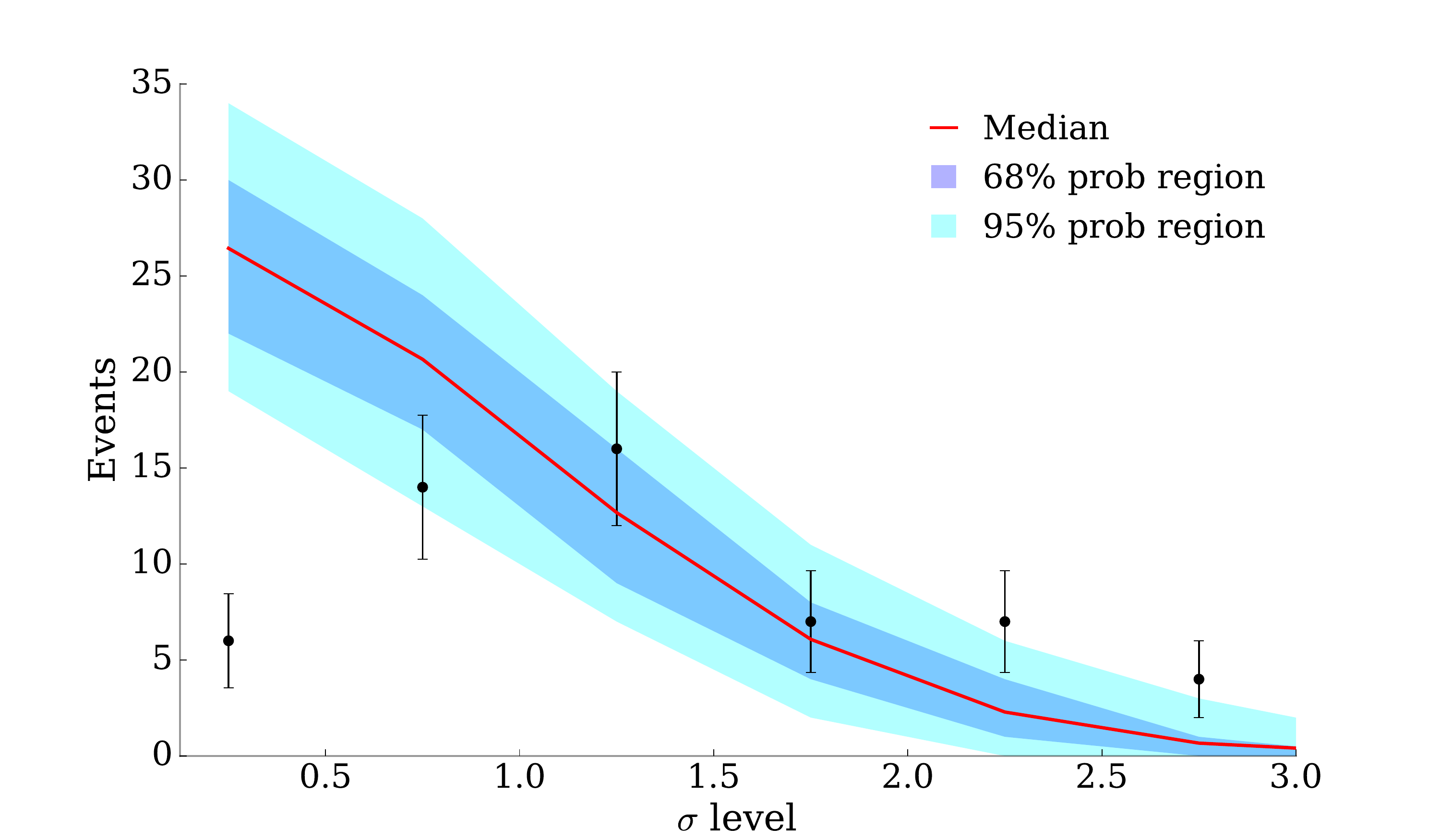}
		\caption{Comparison between the percentile regions for the simulated datasets and all of the DoL data points (69 GRBs).}
		\label{fig:mc_datasets_gbm}
	\end{figure}

	To evaluate if the discrepancy observed in the CDFs is significant or not, a method able to naturally include the statistical noise is needed. We chose to make use of simulations. Under the assumption of absence of systematics (i.e. standard normal distribution), we draw $10^5$ samples of the same size as the original. The synthetic data is then binned exactly as the real one and in each bin the 68\% and 95\% percentile intervals are computed. These intervals, in the limit of large numbers, define probability regions and allow us to assess how extreme the actual data points are. This is done through a simple graphical check \citep{gelman2003, wilk1968}, as in Fig. \ref{fig:mc_datasets}. As one can see, not all points fall inside the 95\% probability region. This suggests that most likely small systematics are still present, albeit to a lesser extent than compared to DoL, which shows a much larger discrepancy (Fig. \ref{fig:mc_datasets_gbm}).
	This DoL discrepancy was \replaced{partially solved}{described} in \citet{2015ApJS..216...32C} by convolving the statistical uncertainties with a second, purely empirical, distribution for the systematic error, i.e. ``The distribution of systematic uncertainties is well represented ($68\%$ confidence
	level) by a 3\fdg7 Gaussian with a non-Gaussian tail that contains about $10\%$ of
	GBM-detected GRBs and extends to approximately $14 \degs$.'' \citep{2015ApJS..216...32C}.
	The uncertainties are thus made larger and the localizations become less precise (larger error region), but more accurate (in 95\% of all GRBs the 2$\sigma$ contour covers the Swift position). 

		Fig. \ref{fig:mc_datasets}
          and Fig. \ref{fig:mc_datasets_gbm} show that, while BALROG
          is able to significantly improve the accuracy of the
          localizations, there appears to be some residual systematics, 
          the origin of which is not yet clear. A significant fraction
          of the inaccurate locations seem to be happen for bursts
          coming from the sides of the spacecraft, where solar panels
          and radiators are situated (near $90/270 \degs$ in
          Fig. \ref{fig:polar_azimuth_offset_sets}). This hints once again 
          to the presence of an inaccurate spacecraft scattering model.

	\subsection{Systematics and error regions}
	\label{subsec:error_contours}
	
\begin{figure}[t]
		\hspace{-0.6cm} \includegraphics[width=0.54\textwidth]{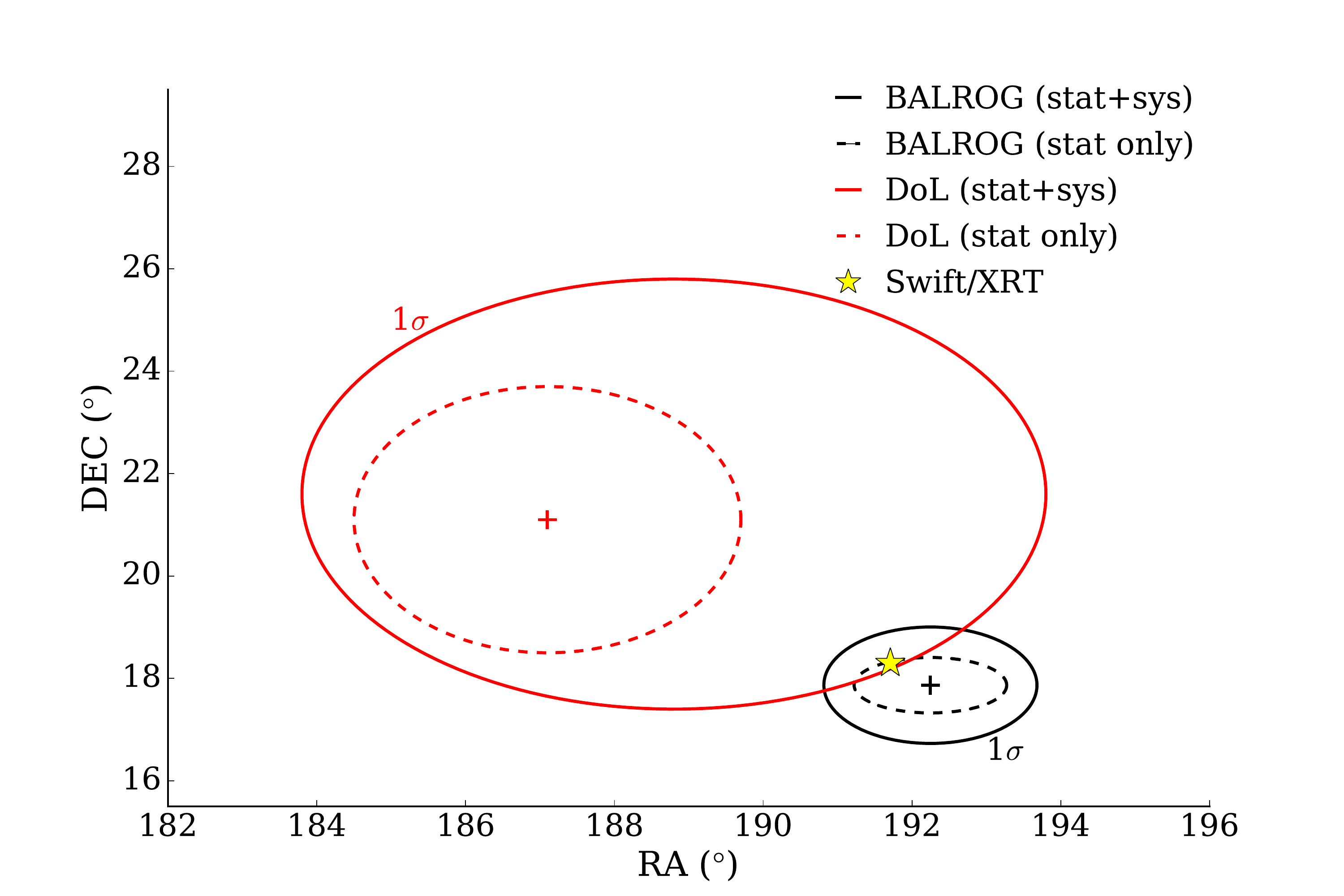}
		\caption{An example of how DoL systematic uncertainties substantially increase the size of the final error regions to achieve accuracy (GRB 170705.115). \added{Here BALROG systematic error is $\sim$1\degs due to the location of the burst in spacecraft coordinates (b0/b1 sides), however this is not the case for all GRBs (up to $\sim$2\degs systematics).}}
		\label{fig:systematic_contour}
	\end{figure}

	One last, but fundamental point for the comparison of the two
        methods is the size of the error regions. This is important in 
        follow-up observations of GBM localizations, if no better position 
        is available from another instrument. If we compare the
        purely statistical uncertainties for the two algorithms, DoL
        achieves smaller errors due to the lack of a real fit of the spectral component. However, these errors lack both
        statistical rigor and the posterior density from the spectral
        model. As described in \cite{2015ApJS..216...32C}, DoL is heavily affected by
        systematics, whereas BALROG is to a much lesser extent. As
        such, the comparison has to be made between DoL
        statistical+systematic uncertainty and BALROG's statistical+systematic
        uncertainty. Since systematics act in the same
        way regardless of how bright the burst is, this implies that
        bright GRBs are the most affected  
        (see example in Fig. \ref{fig:systematic_contour}).
        Due to this, DoL uncertainties end up being much larger than
        BALROG's for bright GRBs, where the contribution from the
        systematics is often dominating the statistical one. 
        Since for a given GBM detection without an external accurate (e.g. \textit{Swift} or other)
        position it cannot be decided whether this GRB is well-localized, this
        convolution with a systematic error has to be applied to all GRBs,
        also those which are well-localized.
		\deleted{Typically, brighter GRBs are more valuable (better spectral constraints, 
        brighter afterglows), so the systematic error matters indeed.}

        A large error region
        is also problematic when dealing with e.g. neutrino coincidence searches,
        where the significance of any potential signal is inversely proportional
        to the size of the search region.

\begin{figure}[t]
		\hspace{-0.6cm} \includegraphics[width=0.54\textwidth]{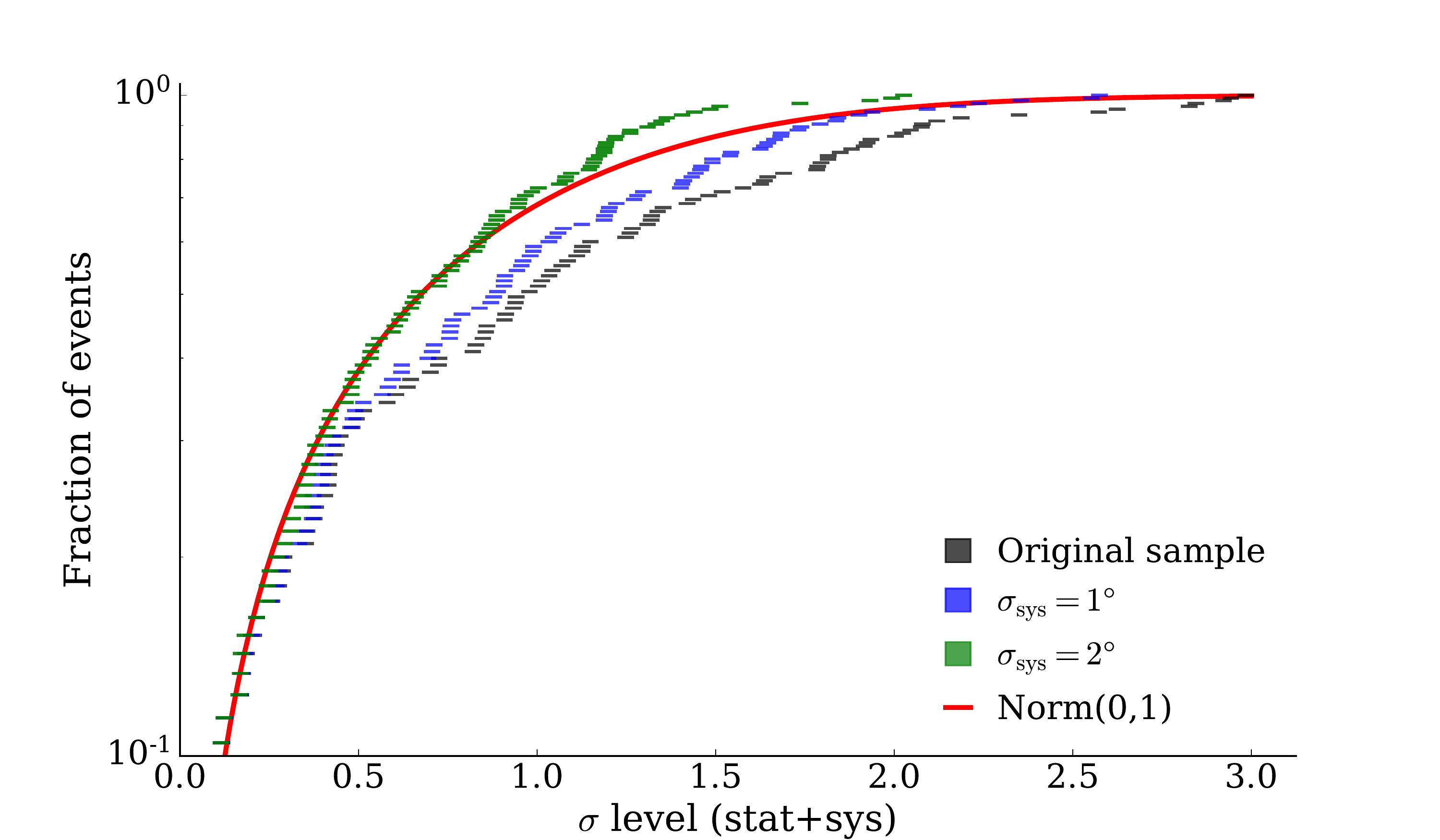}
		\caption{Changes in the BALROG CDF behavior for the full sample as a systematic error is added in quadrature to the statistical one. The y axis is in log scale.}
		\label{fig:sys_cdf}
	\end{figure}

	In sect. \ref{subsec:mc_sim} we have shown that there is evidence for remaining weak systematics. While strictly speaking it is not possible to properly correct for any systematics without first identifying them (and thus expanding the model for source/instrument to naturally include them), we are at the very least able to provide a rough estimate of their magnitude. For the sake of a direct and intuitive comparison with the systematics in DoL, we will adopt the simple procedure of the sum in quadrature to correct for the discrepancy in the CDFs. Based on the fact that most of the inaccurate locations seem to come from the solar panels sides of the spacecraft, we estimate two different systematic errors: one for the GRBs coming from the b0/b1 sides of the spacecraft and one for bursts coming from the solar panel sides. For the full sample, we make the more conservative choice of adopting the larger of the two systematics. Fig. \ref{fig:sys_cdf} shows how adding a systematic error in quadrature can readjust the sigma levels to roughly follow the expected CDF. In particular, we found that the appropriate systematic errors are 1\degs\ for the GRBs detected on the b0/b1 sides and 2\degs\ for the ones detected on the solar panel sides. 
	We stress that this procedure is not meant to properly correct for the remaining systematics, but it is instead only meant as a rough estimate which allows us to compare BALROG error regions to DoL's.

	A simple estimate of the average decrease in sky area to be searched by follow-up telescopes, if BALROG were used instead of DoL, just needs the systematic uncertainties of DoL and BALROG respectively, which need to be added to the statistical localization error in order to obtain a sky region at a certain probability level (e.g. 1, 2 or 3$\sigma$). 
	Let us compute the number of a 2 sigma localization contour.
	We assume the “all-sky” systematic distribution from \cite{2015ApJS..216...32C}, composed of a mixture of two Fisher distributions values of 90\% fraction within 3\fdg7 and a 10\% fraction within 14\fdg3. Integrating the Fisher function to the 95\% containment level implies a $\mathrm{2 \sigma_{sys}} \sim 16\degs$\ median systematic error (see also Figs. 8 and 11 in \cite{2015ApJS..216...32C}). This implies that only GRBs with a statistical error larger than $16\degs$ are not affected by the DoL systematic uncertainty anymore (to more than 10\%). Taking the 2439 GRBs in the present online catalog, only 97 GRBs have a statistical error of 16\degs\ or larger, and thus are unaffected. Thus, 96\% of all GRBs will need an inflated error region being searched. In DoL's best case scenario (localization wise), when the statistical error is very small, the systematical uncertainty will dominate, inflating the 2 sigma error region by about $\mathrm{\pi (2 \sigma_{sys})^2} \sim 800 \deg^2$, compared to BALROG's $\mathrm{\pi (2 \sigma_{sys (BAL)})^2 \sim 50\deg^2}$.
	\newline
	
      Thus, BALROG can not only
	improve substantially the chances of successfully detecting
	any potential multi-wavelength and/or observing multi-messenger
	counterpart, but also save a lot of effort and remove mis-identification potential.
       All this  provides a significant advantage over the DoL algorithm.
	\newline	
	\added{On the other hand,
		the $\sim$30 expected min latency of BALROG positions (10 min until data availability
		at HEASARC, 10-15 min processing time, 5 min human quality check)
		implies a later start of the observations of the fading afterglow,
		negatively impacting the discovery potential. Quantifying the various
		effects is beyond the scope of this paper, but some basic arguments
		are as follows (where we use the PTF follow-up by \citet{2015ApJ...806...52S}
		as an example):
		First, what is the impact of a 30 min delay on the observations?
		Among the 35 GRBs observed by \citet{2015ApJ...806...52S}, only 4 were observed
		within the first hour after the GRB; the rest was observed between
		1-23 hrs, with a mean of 7 hrs (see Tab. 7.2 in \citet{2015PhDT.........6S}).
		For two of these four GRBs, an afterglow
		was found. In both cases, the afterglow was bright, and would have also
		been easily found when starting the observations 30 min later \citep{2015PhDT.........6S},
		thus there would have been no negative impact.
		Second, what is the impact of a smaller localization area? For the
		35 GRBs observed by \citet{2015ApJ...806...52S}, a 6\degr systematic error
		(computed as the weighted mean of the ``core+tail'' error distribution)
		was added in quadrature to the statistical error. This leads to an
		error region which was covered by typically 10 PTF pointings to only
		30--60\%, with a mean of 40\% (see Tab. 7.2 in \citet{2015PhDT.........6S}).
		With the substantially smaller BALROG positions, the full error region
		would have been covered with even less than the 10 PTF pointings,
		implying that 2.5$\times$ more optical afterglows could have been
		discovered.
		On a more general note, covering the large DoL error boxes (statistical
		plus systematic) requires observing times which in the best cases is of
		the order of our 30 min latency; in normal cases even much larger.
		In balance, the improved BALROG positions should allow a substantial
		increase in afterglow discovery rate.}

	\section{Localizing GBM-detected GRBs with BALROG}
	
	For future searches for coincident detections of gravitational
        wave sources with LIGO/Virgo and high-energy transients,
        Fermi-GBM is expected to play a major role. Thus, it seems
        worthwhile to suggest some guidelines to localize Fermi-GBM
        bursts with BALROG\footnote{The BALROG code is available at \url{https://github.com/mpe-grb}} \added{in order to make it successful for every user}. The following rules are not meant to be
        rigidly applied to every GRB, since from a detection
        standpoint there are no two identical bursts. Every event is
        different: not only its spectrum and lightcurve, but also the
        conditions in which the detection happens (detector pointings,
        occultation, background etc.) are never the same. For these
        reasons it is not possible to define strict rules to localize
        GRBs with BALROG. There are, however, some general guidelines
        which work well in the vast majority of cases.  The
        considerations made in this section, though derived purely on
        the bright burst sample, are valid both for bright and faint
        bursts.
	
	\begin{enumerate}	
	\item Time-intervals shorter than 8--10 s: It is good practice
          to avoid selecting excessively long time intervals for the
          source. While a wider interval will include more counts and
          thus improve the photon statistics, this comes at the price
          of a less accurate response, since only a single response is
          computed at the center of the selected time interval. Due to
          its orbit, the boresight of Fermi (and thus that of most
          detectors) slews over 1\degs\ on the sky every 16 s. Thus,
          the wider the interval, in particular if longer than 16 s,
          the less accurate the overall response will be. If possible,
          it is better to try to select a duration with
          $\delta t \lesssim 8-10$ s in order to keep the detector
          response `smearing' below 0\fdg5. At this timescale,
          any inaccuracy in the generated response is negligible
          compared to the statistical uncertainties.
		 
		In addition to the above, spectral evolution may also be an issue. If spectral evolution is significant during the selected time interval, very unreliable locations will be generated, as a good match of the spectrum is an integral part of a good localization result.
	
		To avoid running into this kind of issue, it is best to only select the single most prominent peak in the light curve. In principle it is also possible to fit multiple spectral components at the same time, but this approach is much more time consuming and may prove to be too slow for a rapid localization.
	
	\item	Detectors within 70--80\degs:
		As for detector selection criteria, a good approach is to choose only the ones with angular separations to the source below 70--80\degs. Depending on the conditions of detection, this is not always possible, in which case a less optimal selection needs to be made. A widely used selection criterion for localization and spectral analysis \citep{2014ApJS..211...13V, 2014ApJS..211...12G, 2016AA...588A.135Y} is to select all detectors with angular separations $<$60\degs, but this is found often to be too restrictive for localization purposes. Since the angular separation needs a source position to be computed beforehand, this suggestion implies an iterative approach: the initial location can either be derived using all detectors, or taken from the flight or ground automatic localizations as distributed via GCN.

	\item Not more than 6--7 detectors:
		Generally, the precision of the localization depends strongly on the number of detectors (with signal) up to $\sim$6--7. Beyond this number, additional detectors do not achieve a significant improvement. They can actually sometimes generate less accurate locations. As a very approximate rule, one should try to select at least four NaI and one BGO detectors, whenever the signal is strong enough to allow for this. For very soft bursts, the BGO detectors will most likely carry no significant signal, and can be excluded.
	
	\item Prefer one detector side: unless GRBs illuminate GBM at a satellite azimuth angle consistent with the solar panels, the selection of detectors from either b0 or b1 seems to provide more accurate localizations than including a detector on the opposite side. \added{One can easily establish from which side of the spacecraft the burst is coming by simply comparing the counts in b0 and b1. This can be done both manually with an inspection of the lightcurves or automated with some code.}

	\item	Use multiple spectral models:
		Finally, regarding spectrum selection, there is no general rule to apply until we fully understand the physical processes shaping the prompt GRB spectra. Ideally, the fits should be made with a particular physical model in mind (e.g. synchrotron, photospheric etc.), however this approach is both very time consuming and poorly applicable for the coarse trigdat data. For the purpose of localizing bursts it seems sufficient to use empirical spectra, such as a cut-off power-law or a Band function but with all parameters left free. 
	\end{enumerate}
		
	\section{Conclusions}

	The performance of localizing Fermi-GBM-detected GRBs with the BALROG
	scheme  \citep{2018MNRAS.476.1427B} has been analyzed. The sample employed, 
	while not very large, is sufficient to show that a significant improvement 
	of the localizations is achieved over the previous approach 
	\citep{2015ApJS..216...32C}, both in terms of precision (i.e. size of 
	error region) and accuracy (i.e. offset from independent localizations 
	with {\it Swift}).
	
	A search for possible remaining systematics was also
        performed. The BALROG scheme with its simultaneous spectral
        and location fitting, as well as the improved statistical
        modeling, removed most of what was previously considered
        instrument systematics. The remaining systematic error using BALROG is \added{approximately} 1--2\degs\,
        as compared to the 3\fdg7 core plus 14\fdg3 tail in \cite{2015ApJS..216...32C}.
        We find that GRBs are located more precisely when arriving on the
        b0/b1 side of the satellite, in which case the BALROG systematic
        error is even lower, only 1\degs. This asymmetry \added{(hints of which were already found in \citet{2015ApJS..216...32C})} is not well understood, and warrants further study which may be able to shed more light
        onto this effect.

        There are two immediate implications beyond the Fermi-GBM
        catalog localizations: First, the DoL localization procedure
        derives from that used for the BATSE instrument on the Compton
        Gamma-ray Observatory, so all BATSE locations of GRBs
        \citep{1999ApJS..122..465P} suffer the same systematic problem
        as the Fermi-GBM DoL locations.  Second, since gamma-ray
        spectra and localizations are intricately connected, the
        prompt gamma-ray spectra of GRBs which are derived with
        DoL-localizations will suffer a systematic deformation. \added{The extent of this effect is not yet known, but it should be carefully considered when employing DoL localizations.} We
        therefore call for caution when using the spectral parameters
        as published in the Fermi GBM spectral catalogs
        \citep{2012ApJS..199...19G, 2014ApJS..211...12G,
          2016AA...588A.135Y}.

	BALROG has superior performance over the previous DoL method,
        and thus should be used for future multi-wavelength and
        multi-messenger studies.  This will guarantee that precious
        observing time at other observatories and facilities is not
        wasted by searching wrong sky areas.

	\section*{Acknowledgements}
	
	We acknowledge technical discussions with A.v. Kienlin during
        an early stage of this work. We are grateful to Daniel Mortlock and Hoi-Fung (David) Yu for inspiring interactions. FB and JG acknowledge support
        provided by the DFG through SFB 1258.
	JMB and JG acknowledge support provided by the DFG cluster of excellence ``Origin and Structure of the Universe''.\\

\begin{deluxetable*}{c c c c c c c c c c c c c c c}
	\caption{List of the GRBs analyzed. In the model column, the abbreviations have the following meanings: cpl $=$ cut-off power-law, band $=$ Band function and pl $=$ power-law. Asterisks in trigger numbers denote short GRBs. GRBs after 130609.902 constitute sample 2).}
	\label{tab:grbs}
	\tablehead{
		\rule{0pt}{3ex}		
		\multirow{3}{*}{Trigger} & \multicolumn{2}{c}{{\it Swift}} & \multicolumn{4}{c}{BALROG} & \multicolumn{4}{c}{DoL} & \multicolumn{2}{c}{Data} 
		\\\cmidrule(lr){2-3}\cmidrule(lr){4-7}\cmidrule(lr){8-11}\cmidrule(lr){12-13}
		\colhead{} & \multirow{2}{*}{RA} & \multirow{2}{*}{DEC} & \multirow{2}{*}{RA} & \multirow{2}{*}{DEC} & \multirow{2}{*}{Offset} & \multirow{2}{*}{Model}  & \multirow{2}{*}{RA}  & \multirow{2}{*}{DEC} & \multirow{2}{*}{Err.} & \multirow{2}{*}{Offset} & \multirow{2}{*}{Fluence} & \multirow{2}{*}{Peak flux}\\
		\colhead{} & \colhead{($\degs$)} & \colhead{($\degs$)}  & \colhead{($\degs$)} & \colhead{($\degs$)} & \colhead{($\degs$)} & \colhead{}  & \colhead{($\degs$)}  & \colhead{($\degs$)} & \colhead{($\degs$)} & \colhead{($\degs$)} & \colhead{($\mathrm{erg \cdot cm^{-2}}$)} & \colhead{($\mathrm{ph \cdot cm^{-2} \cdot s^{-1}}$)}
	}
	\startdata
	080714745 & 188.1 & -60.2 & $193.6 \pm 3.4$ & $-57.9 \pm 2.7$ & 3.7 & cpl & 183.5 & -57.5 & 3.9 & 3.6 & $6.3 \cdot 10^{-6}$ & 6.9 \\
	080723557 & 176.8 & -60.2 & $179.3 \pm 0.5$ & $-60.8 \pm 0.5$ & 1.4 & cpl & 175.1 & -60.7 & 1.0 & 1.0 & $7.2 \cdot 10^{-5}$ & $3.0 \cdot 10^{1}$ \\
	080725435 & 121.7 & -14.0 & $120.7 \pm 1.7$ & $-12.7 \pm 2.2$ & 1.6 & cpl & 123.1 & -23.1 & 2.2 & 9.2 & $8.0 \cdot 10^{-6}$ & 3.4 \\
	080727964 & 32.6 & 64.1 & $41.6 \pm 6.7$ & $63.3 \pm 2.8$ & 4.1 & cpl & 40.0 & 62.2 & 2.7 & 3.8 & $1.3 \cdot 10^{-5}$ & 3.5 \\
	080804972 & 328.7 & -53.2 & $330.0 \pm 3.9$ & $-52.9 \pm 1.4$ & 0.9 & cpl & 320.8 & -52.7 & 2.4 & 4.8 & $9.1 \cdot 10^{-6}$ & 3.8 \\
	080916009 & 119.8 & -56.6 & $120.6 \pm 1.4$ & $-58.5 \pm 0.8$ & 1.9 & band & 124.4 & -54.7 & 1.4 & 3.2 & $6.0 \cdot 10^{-5}$ & $1.4 \cdot 10^{1}$ \\
	081102739 & 331.2 & 53.0 & $326.9 \pm 4.9$ & $50.1 \pm 3.4$ & 3.9 & pl & 321.3 & 51.9 & 4.5 & 6.1 & $3.8 \cdot 10^{-6}$ & 2.7 \\
	081121858 & 89.3 & -60.6 & $94.3 \pm 5.2$ & $-63.9 \pm 1.0$ & 4.0 & cpl & 97.2 & -59.9 & 1.1 & 4.0 & $1.5 \cdot 10^{-5}$ & 7.7 \\
	081126899 & 323.5 & 48.7 & $313.6 \pm 9.1$ & $49.3 \pm 4.2$ & 6.5 & cpl & 326.9 & 50.3 & 2.7 & 2.7 & $1.1 \cdot 10^{-5}$ & 6.5 \\
	081221681 & 15.8 & -24.5 & $14.8 \pm 0.3$ & $-24.2 \pm 0.4$ & 1.0 & cpl & 14.1 & -25.2 & 1.2 & 1.7 & $3.0 \cdot 10^{-5}$ & $2.5 \cdot 10^{1}$ \\
	081222204 & 22.7 & -34.1 & $24.2 \pm 1.3$ & $-33.3 \pm 0.9$ & 1.5 & cpl & 18.6 & -32.4 & 1.5 & 3.8 & $1.2 \cdot 10^{-5}$ & $1.3 \cdot 10^{1}$ \\
	090102122 & 128.2 & 33.1 & $124.6 \pm 0.7$ & $31.5 \pm 0.5$ & 3.4 & cpl & 128.5 & 30.3 & 1.5 & 2.8 & $2.8 \cdot 10^{-5}$ & $1.1 \cdot 10^{1}$ \\
	090129880 & 269.1 & -32.8 & $268.3 \pm 5.1$ & $-33.3 \pm 1.4$ & 0.8 & cpl & 270.6 & -33.8 & 1.8 & 1.6 & $5.6 \cdot 10^{-6}$ & 6.6 \\
	090323002 & 190.7 & 17.1 & $190.3 \pm 2.3$ & $16.2 \pm 2.0$ & 1.0 & cpl & 190.1 & 21.1 & 1.2 & 4.0 & $1.1 \cdot 10^{-4}$ & $1.3 \cdot 10^{1}$ \\
	090328401 & 90.9 & -42.0 & $94.0 \pm 1.6$ & $-44.0 \pm 0.6$ & 3.0 & band & 95.9 & -45.1 & 1.0 & 4.8 & $4.2 \cdot 10^{-5}$ & $1.7 \cdot 10^{1}$ \\
	090424592 & 189.5 & 16.8 & $190.3 \pm 0.2$ & $17.9 \pm 0.3$ & 1.3 & cpl & 191.5 & 18.1 & 1.6 & 2.3 & $4.6 \cdot 10^{-5}$ & $1.1 \cdot 10^{2}$ \\
	090510016* & 333.6 & -26.6 & $334.6 \pm 1.7$ & $-30.3 \pm 1.6$ & 3.8 & cpl & 335.0 & -31.3 & 1.0 & 4.9 & $3.4 \cdot 10^{-6}$ & 9.1 \\
	090531775* & 252.1 & -36.0 & $250.0 \pm 6.3$ & $-33.7 \pm 3.4$ & 2.9 & cpl & 252.9 & -31.5 & 7.2 & 4.5 & $3.2 \cdot 10^{-7}$ & 3.4 \\
	090618353 & 294.0 & 78.4 & $285.5 \pm 3.0$ & $79.7 \pm 0.3$ & 2.0 & cpl & 288.4 & 80.0 & 1.0 & 1.9 & $2.7 \cdot 10^{-4}$ & $6.9 \cdot 10^{1}$ \\
	090813174 & 225.1 & 88.6 & $282.0 \pm 32.8$ & $87.5 \pm 1.2$ & 2.1 & cpl & 40.2 & 86.0 & 5.1 & 5.4 & $3.3 \cdot 10^{-6}$ & $1.4 \cdot 10^{1}$ \\
	090902462 & 264.9 & 27.3 & $263.3 \pm 0.3$ & $27.8 \pm 0.3$ & 1.5 & cpl & 261.4 & 26.1 & 1.0 & 3.3 & $2.2 \cdot 10^{-4}$ & $7.7 \cdot 10^{1}$ \\
	090904058 & 264.2 & -25.2 & $260.5 \pm 1.5$ & $-23.5 \pm 2.5$ & 3.8 & cpl & 265.9 & -30.1 & 1.9 & 5.1 & $2.2 \cdot 10^{-5}$ & 6.8 \\
	090926181 & 353.4 & -66.3 & $353.7 \pm 0.6$ & $-66.2 \pm 0.2$ & 0.2 & cpl & 350.1 & -63.5 & 1.0 & 3.1 & $1.5 \cdot 10^{-4}$ & $8.1 \cdot 10^{1}$ \\
	091003191 & 251.5 & 36.6 & $251.2 \pm 0.3$ & $36.3 \pm 0.2$ & 0.4 & cpl & 251.2 & 37.3 & 1.0 & 0.7 & $2.3 \cdot 10^{-5}$ & $2.9 \cdot 10^{1}$ \\
	091020900 & 175.7 & 51.0 & $180.8 \pm 4.2$ & $52.4 \pm 2.5$ & 3.5 & cpl & 174.4 & 52.7 & 3.1 & 1.9 & $8.3 \cdot 10^{-6}$ & 6.8 \\
	091112737 & 257.7 & -36.7 & $258.2 \pm 2.0$ & $-38.9 \pm 5.0$ & 2.2 & cpl & 258.3 & -36.7 & 3.2 & 0.5 & $9.9 \cdot 10^{-6}$ & 4.2 \\
	091127976 & 36.6 & -19.1 & $35.9 \pm 0.5$ & $-18.4 \pm 0.4$ & 0.9 & cpl & 38.1 & -21.0 & 1.1 & 2.4 & $2.1 \cdot 10^{-5}$ & $6.8 \cdot 10^{1}$ \\
	091208410 & 29.4 & 16.9 & $28.5 \pm 0.9$ & $16.7 \pm 1.1$ & 0.9 & pl & 30.1 & 13.5 & 2.6 & 3.5 & $6.2 \cdot 10^{-6}$ & $2.1 \cdot 10^{1}$ \\
	091221870 & 55.8 & 23.2 & $57.4 \pm 2.4$ & $22.2 \pm 1.4$ & 1.8 & cpl & 54.5 & 27.3 & 1.2 & 4.3 & $8.9 \cdot 10^{-6}$ & 4.3 \\
	100401297 & 290.8 & -8.3 & $292.2 \pm 9.2$ & $-15.5 \pm 7.7$ & 7.4 & pl & 290.0 & -16.3 & 7.9 & 8.0 & $1.9 \cdot 10^{-6}$ & 4.1 \\
	100414097 & 192.1 & 8.7 & $191.1 \pm 0.4$ & $9.3 \pm 0.3$ & 1.2 & band & 185.7 & 15.7 & 1.0 & 9.4 & $8.8 \cdot 10^{-5}$ & $2.2 \cdot 10^{1}$ \\
	100427356 & 89.2 & -3.5 & $85.7 \pm 1.3$ & $-2.0 \pm 2.1$ & 3.8 & cpl & 91.0 & -1.4 & 2.6 & 2.8 & $2.3 \cdot 10^{-6}$ & 3.8 \\
	100522157 & 7.0 & 9.5 & $9.5 \pm 1.4$ & $11.5 \pm 1.2$ & 3.2 & cpl & 8.0 & 10.5 & 3.9 & 1.4 & $3.9 \cdot 10^{-6}$ & $1.1 \cdot 10^{1}$ \\
	100615083 & 177.2 & -19.5 & $173.9 \pm 2.3$ & $-18.9 \pm 2.1$ & 3.2 & pl & 175.6 & -21.1 & 2.4 & 2.2 & $8.7 \cdot 10^{-6}$ & 8.3 \\
	100619015 & 84.6 & -27.1 & $86.3 \pm 3.3$ & $-24.6 \pm 3.3$ & 2.9 & pl & 84.0 & -25.5 & 6.6 & 1.7 & $1.1 \cdot 10^{-5}$ & 7.4 \\
	\enddata
\end{deluxetable*}

\begin{deluxetable*}{c c c c c c c c c c c c c c c}
	\tablehead{
		\rule{0pt}{3ex}		
		\multirow{3}{*}{Trigger} & \multicolumn{2}{c}{{\it Swift}} & \multicolumn{4}{c}{BALROG} & \multicolumn{4}{c}{DoL} & \multicolumn{2}{c}{Data} 
		\\\cmidrule(lr){2-3}\cmidrule(lr){4-7}\cmidrule(lr){8-11}\cmidrule(lr){12-13}
		\colhead{} & \multirow{2}{*}{RA} & \multirow{2}{*}{DEC} & \multirow{2}{*}{RA} & \multirow{2}{*}{DEC} & \multirow{2}{*}{Offset} & \multirow{2}{*}{Model}  & \multirow{2}{*}{RA}  & \multirow{2}{*}{DEC} & \multirow{2}{*}{Err.} & \multirow{2}{*}{Offset} & \multirow{2}{*}{Fluence} & \multirow{2}{*}{Peak flux}\\
		\colhead{} & \colhead{($\degs$)} & \colhead{($\degs$)}  & \colhead{($\degs$)} & \colhead{($\degs$)} & \colhead{($\degs$)} & \colhead{}  & \colhead{($\degs$)}  & \colhead{($\degs$)} & \colhead{($\degs$)} & \colhead{($\degs$)} & \colhead{($\mathrm{erg \cdot cm^{-2}}$)} & \colhead{($\mathrm{ph \cdot cm^{-2} \cdot s^{-1}}$)}
	}
	\startdata
	100704149 & 133.6 & -24.2 & $133.4 \pm 2.2$ & $-22.3 \pm 4.0$ & 1.9 & cpl & 133.2 & -23.6 & 1.6 & 0.7 & $8.4 \cdot 10^{-6}$ & 7.2 \\
	100728095 & 88.8 & -15.3 & $91.8 \pm 1.0$ & $-14.2 \pm 0.9$ & 3.1 & cpl & 88.3 & -13.7 & 1.0 & 1.7 & $1.3 \cdot 10^{-4}$ & $1.1 \cdot 10^{1}$ \\
	100728439 & 44.1 & 0.3 & $43.9 \pm 2.8$ & $1.3 \pm 1.9$ & 1.0 & pl & 41.5 & 0.2 & 4.2 & 2.6 & $3.3 \cdot 10^{-6}$ & 6.2 \\
	100906576 & 28.7 & 55.6 & $31.0 \pm 0.9$ & $54.5 \pm 0.6$ & 1.7 & cpl & 28.0 & 55.2 & 1.1 & 0.6 & $2.3 \cdot 10^{-5}$ & $1.4 \cdot 10^{1}$ \\
	101023951 & 318.0 & -65.5 & $322.5 \pm 1.5$ & $-68.1 \pm 0.4$ & 3.2 & cpl & 315.5 & -66.5 & 1.0 & 1.4 & $6.4 \cdot 10^{-5}$ & $3.7 \cdot 10^{1}$ \\
	101024486 & 66.5 & -77.3 & $96.6 \pm 53.0$ & $-75.8 \pm 4.6$ & 7.1 & pl & 147.1 & -77.2 & 9.6 & 16.4 & $3.3 \cdot 10^{-6}$ & 8.3 \\
	101201418 & 1.9 & -16.1 & $3.0 \pm 0.8$ & $-14.1 \pm 1.2$ & 2.3 & band & 3.9 & -14.7 & 1.6 & 2.4 & $2.4 \cdot 10^{-5}$ & 6.9 \\
	110102788 & 245.9 & 7.6 & $243.6 \pm 1.0$ & $5.4 \pm 0.9$ & 3.2 & cpl & 246.2 & 6.0 & 2.0 & 1.6 & $3.7 \cdot 10^{-5}$ & $1.4 \cdot 10^{1}$ \\
	110213220 & 43.0 & 49.3 & $39.5 \pm 1.7$ & $47.9 \pm 1.2$ & 2.7 & pl & 49.0 & 52.8 & 2.3 & 5.1 & $9.4 \cdot 10^{-6}$ & $1.8 \cdot 10^{1}$ \\
	110318552 & 338.3 & -15.3 & $337.9 \pm 0.9$ & $-17.8 \pm 1.1$ & 2.5 & cpl & 335.9 & -14.9 & 1.9 & 2.4 & $8.2 \cdot 10^{-6}$ & $1.1 \cdot 10^{1}$ \\
	110402009 & 197.4 & 61.4 & $196.8 \pm 9.6$ & $58.0 \pm 3.8$ & 3.3 & cpl & 187.7 & 58.7 & 2.2 & 5.5 & $1.1 \cdot 10^{-5}$ & 7.8 \\
	110610640 & 308.2 & 74.8 & $301.0 \pm 5.7$ & $73.6 \pm 1.3$ & 2.3 & pl & 306.5 & 75.9 & 2.6 & 1.2 & $8.0 \cdot 10^{-6}$ & 5.8 \\
	110625881 & 286.8 & 6.8 & $287.6 \pm 0.2$ & $7.9 \pm 0.2$ & 1.4 & cpl & 287.7 & 6.7 & 1.0 & 0.9 & $6.5 \cdot 10^{-5}$ & $7.7 \cdot 10^{1}$ \\
	110709642 & 238.9 & 40.9 & $238.7 \pm 1.4$ & $38.8 \pm 1.6$ & 2.1 & cpl & 241.2 & 41.8 & 1.1 & 1.9 & $3.7 \cdot 10^{-5}$ & $1.1 \cdot 10^{1}$ \\
	110731465 & 280.5 & -28.5 & $282.4 \pm 0.4$ & $-28.8 \pm 0.9$ & 1.7 & cpl & 283.1 & -34.0 & 1.0 & 5.9 & $2.3 \cdot 10^{-5}$ & $2.1 \cdot 10^{1}$ \\
	111228657 & 150.1 & 18.3 & $148.6 \pm 0.4$ & $17.8 \pm 1.3$ & 1.5 & pl & 146.6 & 14.6 & 2.4 & 5.0 & $1.8 \cdot 10^{-5}$ & $2.1 \cdot 10^{1}$ \\
	120102095 & 276.2 & 24.7 & $276.2 \pm 2.1$ & $22.0 \pm 0.8$ & 2.7 & cpl & 277.1 & 20.4 & 2.0 & 4.4 & $1.3 \cdot 10^{-5}$ & $1.9 \cdot 10^{1}$ \\
	120119170 & 120.0 & -9.1 & $120.1 \pm 0.5$ & $-9.0 \pm 0.6$ & 0.1 & cpl & 119.0 & -9.0 & 1.1 & 1.0 & $3.9 \cdot 10^{-5}$ & $1.7 \cdot 10^{1}$ \\
	120326056 & 273.9 & 69.3 & $267.3 \pm 5.7$ & $71.6 \pm 2.0$ & 3.2 & pl & 262.2 & 62.1 & 4.2 & 8.6 & $3.3 \cdot 10^{-6}$ & 7.7 \\
	120624933 & 170.9 & 8.9 & $170.8 \pm 0.9$ & $10.8 \pm 0.8$ & 1.9 & cpl & 171.8 & 5.6 & 1.0 & 3.4 & $1.9 \cdot 10^{-4}$ & $1.8 \cdot 10^{1}$ \\
	120703726 & 339.4 & -29.7 & $335.6 \pm 1.5$ & $-30.3 \pm 1.2$ & 3.3 & cpl & 339.8 & -29.5 & 1.7 & 0.4 & $8.3 \cdot 10^{-6}$ & $1.7 \cdot 10^{1}$ \\
	120913997 & 213.6 & -14.5 & $211.7 \pm 2.7$ & $-18.7 \pm 1.6$ & 4.6 & pl & 214.8 & -16.6 & 1.5 & 2.4 & $2.0 \cdot 10^{-5}$ & 5.3 \\
	121031949 & 170.8 & -3.5 & $167.7 \pm 2.5$ & $-5.3 \pm 2.9$ & 3.6 & pl & 173.1 & -1.9 & 3.4 & 2.8 & $1.5 \cdot 10^{-5}$ & 7.4 \\
	121128212 & 300.6 & 54.3 & $300.3 \pm 1.9$ & $57.3 \pm 1.4$ & 3.0 & cpl & 278.8 & 41.6 & 1.5 & 19.2 & $9.3 \cdot 10^{-6}$ & $1.8 \cdot 10^{1}$ \\
	130216790 & 58.9 & 2.0 & $61.6 \pm 1.6$ & $-3.2 \pm 1.7$ & 5.9 & pl & 61.7 & 3.5 & 2.3 & 3.2 & $4.9 \cdot 10^{-6}$ & $1.5 \cdot 10^{1}$ \\
	130216927 & 67.9 & 14.7 & $70.8 \pm 0.7$ & $14.3 \pm 1.3$ & 2.8 & pl & 69.4 & 16.3 & 1.5 & 2.2 & $5.9 \cdot 10^{-6}$ & 9.2 \\
	130305486 & 116.8 & 52.0 & $115.5 \pm 0.6$ & $51.6 \pm 0.6$ & 0.9 & cpl & 119.7 & 49.0 & 1.0 & 3.5 & $4.6 \cdot 10^{-5}$ & $2.7 \cdot 10^{1}$ \\
	130306991 & 279.5 & -11.8 & $278.1 \pm 0.6$ & $-10.4 \pm 0.4$ & 1.9 & cpl & 276.9 & -11.5 & 1.0 & 2.6 & $1.2 \cdot 10^{-4}$ & $2.9 \cdot 10^{1}$ \\
	130420313 & 196.1 & 59.4 & $198.7 \pm 22.7$ & $59.6 \pm 5.2$ & 1.3 & cpl & 205.7 & 58.8 & 4.4 & 5.0 & $1.2 \cdot 10^{-5}$ & 5.4 \\
	130427324 & 173.1 & 27.7 & $172.3 \pm 0.5$ & $28.5 \pm 0.5$ & 1.0 & cpl & 172.5 & 25.5 & 1.0 & 2.3 & $2.5 \cdot 10^{-3}$ & $1.1 \cdot 10^{3}$ \\
	130502327 & 66.8 & 71.1 & $67.1 \pm 1.9$ & $71.0 \pm 0.6$ & 0.1 & cpl & 77.0 & 70.3 & 1.0 & 3.5 & $1.0 \cdot 10^{-4}$ & $4.6 \cdot 10^{1}$ \\
	130504978 & 91.6 & 3.8 & $93.8 \pm 0.5$ & $4.2 \pm 0.3$ & 2.2 & cpl & 90.7 & 4.3 & 1.0 & 1.0 & $1.3 \cdot 10^{-4}$ & $4.3 \cdot 10^{1}$ \\
	130518580 & 355.7 & 47.5 & $352.8 \pm 0.3$ & $46.9 \pm 0.3$ & 2.1 & cpl & 356.3 & 47.0 & 1.0 & 0.6 & $9.5 \cdot 10^{-5}$ & $4.5 \cdot 10^{1}$ \\
	130609902 & 53.8 & -40.2 & $50.1 \pm 1.7$ & $-37.9 \pm 1.3$ & 3.7 & cpl & 51.9 & -42.9 & 1.0 & 3.1 & $5.4 \cdot 10^{-5}$ & $1.4 \cdot 10^{1}$ \\
	\enddata
\end{deluxetable*}

\begin{deluxetable*}{c c c c c c c c c c c c c c c}
	\tablehead{
		\rule{0pt}{3ex}		
		\multirow{3}{*}{Trigger} & \multicolumn{2}{c}{{\it Swift}} & \multicolumn{4}{c}{BALROG} & \multicolumn{4}{c}{DoL} & \multicolumn{2}{c}{Data} 
		\\\cmidrule(lr){2-3}\cmidrule(lr){4-7}\cmidrule(lr){8-11}\cmidrule(lr){12-13}
		\colhead{} & \multirow{2}{*}{RA} & \multirow{2}{*}{DEC} & \multirow{2}{*}{RA} & \multirow{2}{*}{DEC} & \multirow{2}{*}{Offset} & \multirow{2}{*}{Model}  & \multirow{2}{*}{RA}  & \multirow{2}{*}{DEC} & \multirow{2}{*}{Err.} & \multirow{2}{*}{Offset} & \multirow{2}{*}{Fluence} & \multirow{2}{*}{Peak flux}\\
		\colhead{} & \colhead{($\degs$)} & \colhead{($\degs$)}  & \colhead{($\degs$)} & \colhead{($\degs$)} & \colhead{($\degs$)} & \colhead{}  & \colhead{($\degs$)}  & \colhead{($\degs$)} & \colhead{($\degs$)} & \colhead{($\degs$)} & \colhead{($\mathrm{erg \cdot cm^{-2}}$)} & \colhead{($\mathrm{ph \cdot cm^{-2} \cdot s^{-1}}$)}
	}
	\startdata
	130727698 & 330.8 & -65.5 & $328.6 \pm 4.0$ & $-63.3 \pm 6.4$ & 2.4 & cpl & 326.1 & -64.1 & 2.5 & 2.4 & $8.2 \cdot 10^{-6}$ & $1.1 \cdot 10^{1}$ \\
	130925173 & 41.2 & -26.1 & $33.9 \pm 3.4$ & $-27.9 \pm 2.3$ & 6.8 & pl & - & - & - & - & $8.5 \cdot 10^{-5}$ & $1.1 \cdot 10^{1}$ \\
	131229277 & 85.2 & -4.4 & $86.5 \pm 1.1$ & $-6.8 \pm 1.3$ & 2.7 & cpl & - & - & - & - & $2.6 \cdot 10^{-5}$ & $2.4 \cdot 10^{1}$ \\
	140108721 & 325.1 & 58.7 & $326.2 \pm 2.3$ & $59.3 \pm 1.2$ & 0.8 & cpl & - & - & - & - & $2.0 \cdot 10^{-5}$ & $1.0 \cdot 10^{1}$ \\
	140206304 & 145.3 & 66.8 & $147.9 \pm 1.7$ & $68.3 \pm 1.2$ & 1.8 & cpl & - & - & - & - & $1.6 \cdot 10^{-5}$ & $1.7 \cdot 10^{1}$ \\
	140209313* & 81.3 & 32.5 & $84.2 \pm 0.2$ & $31.2 \pm 0.3$ & 2.8 & cpl & - & - & - & - & $9.0 \cdot 10^{-6}$ & $5.8 \cdot 10^{1}$ \\
	140213807 & 105.2 & -73.1 & $103.7 \pm 16.9$ & $-72.0 \pm 1.1$ & 1.3 & cpl & - & - & - & - & $2.1 \cdot 10^{-5}$ & $3.7 \cdot 10^{1}$ \\
	140323433 & 357.0 & -79.9 & $357.6 \pm 4.7$ & $-76.7 \pm 1.2$ & 3.2 & cpl & - & - & - & - & $3.2 \cdot 10^{-5}$ & 9.7 \\
	140506880 & 276.8 & -55.6 & $275.5 \pm 5.7$ & $-56.3 \pm 3.0$ & 1.0 & cpl & - & - & - & - & $6.6 \cdot 10^{-6}$ & $1.6 \cdot 10^{1}$ \\
	140512814 & 289.4 & -15.1 & $289.9 \pm 3.5$ & $-15.8 \pm 2.9$ & 0.9 & cpl & - & - & - & - & $2.9 \cdot 10^{-5}$ & $1.1 \cdot 10^{1}$ \\
	140716436 & 108.1 & -60.1 & $102.1 \pm 2.7$ & $-65.2 \pm 2.2$ & 5.8 & cpl & 109.8 & -61.5 & 2.2 & 1.6 & $1.2 \cdot 10^{-5}$ & $1.1 \cdot 10^{1}$ \\
	141004973 & 76.7 & 12.8 & $77.2 \pm 2.6$ & $11.6 \pm 3.1$ & 1.3 & cpl & - & - & - & - & $1.2 \cdot 10^{-6}$ & 9.8 \\
	141220252 & 195.1 & 32.1 & $197.3 \pm 0.9$ & $33.3 \pm 1.5$ & 2.2 & pl & - & - & - & - & $5.3 \cdot 10^{-6}$ & $1.2 \cdot 10^{1}$ \\
	150201574 & 11.8 & -37.6 & $12.0 \pm 0.6$ & $-37.7 \pm 0.3$ & 0.1 & cpl & - & - & - & - & $6.3 \cdot 10^{-5}$ & $8.9 \cdot 10^{1}$ \\
	150309958 & 277.0 & 86.4 & $276.8 \pm 15.2$ & $86.6 \pm 0.9$ & 0.2 & cpl & 284.3 & 83.3 & 1.0 & 3.2 & $4.0 \cdot 10^{-5}$ & $1.5 \cdot 10^{1}$ \\
	150323395 & 260.5 & 38.3 & $262.8 \pm 3.5$ & $36.4 \pm 2.9$ & 2.7 & cpl & 262.1 & 37.6 & 1.7 & 1.5 & $1.9 \cdot 10^{-5}$ & 8.9 \\
	150403913 & 311.5 & -62.7 & $310.1 \pm 0.7$ & $-60.5 \pm 0.5$ & 2.3 & cpl & 313.5 & -61.0 & 1.0 & 1.9 & $5.5 \cdot 10^{-5}$ & $3.3 \cdot 10^{1}$ \\
	150430015 & 326.5 & -27.9 & $325.2 \pm 1.1$ & $-27.4 \pm 1.7$ & 1.3 & cpl & 324.6 & -25.3 & 1.6 & 3.1 & $1.6 \cdot 10^{-5}$ & 9.9 \\
	150817087 & 249.6 & -12.1 & $251.0 \pm 1.5$ & $-11.7 \pm 1.4$ & 1.4 & cpl & - & - & - & - & $1.2 \cdot 10^{-5}$ & $1.6 \cdot 10^{1}$ \\
	151027166 & 272.5 & 61.4 & $269.3 \pm 3.6$ & $62.1 \pm 0.6$ & 1.7 & cpl & 269.8 & 61.4 & 1.1 & 1.3 & $1.4 \cdot 10^{-5}$ & $1.1 \cdot 10^{1}$ \\
	151229285 & 329.4 & -20.7 & $326.5 \pm 2.2$ & $-20.4 \pm 1.7$ & 2.7 & pl & - & - & - & - & $1.1 \cdot 10^{-6}$ & $1.1 \cdot 10^{1}$ \\
	160325291 & 15.7 & -72.7 & $13.4 \pm 3.4$ & $-74.0 \pm 1.0$ & 1.5 & cpl & - & - & - & - & $1.9 \cdot 10^{-5}$ & 8.5 \\
	160905471 & 162.2 & -50.8 & $159.8 \pm 0.5$ & $-53.5 \pm 0.3$ & 3.1 & cpl & - & - & - & - & $7.3 \cdot 10^{-5}$ & $1.6 \cdot 10^{1}$ \\
	161004964 & 112.2 & -39.9 & $110.0 \pm 0.5$ & $-37.7 \pm 0.7$ & 2.8 & band & - & - & - & - & $1.7 \cdot 10^{-5}$ & $1.6 \cdot 10^{1}$ \\
	161117066 & 322.1 & -29.6 & $323.8 \pm 0.8$ & $-28.3 \pm 0.7$ & 2.0 & cpl & - & - & - & - & $3.1 \cdot 10^{-5}$ & $1.0 \cdot 10^{1}$ \\
	170126480 & 263.6 & -64.8 & $264.2 \pm 2.4$ & $-67.7 \pm 1.1$ & 3.0 & cpl & - & - & - & - & $8.5 \cdot 10^{-6}$ & $1.2 \cdot 10^{1}$ \\
	170208940 & 127.1 & -9.0 & $127.6 \pm 2.0$ & $-10.4 \pm 2.4$ & 1.5 & cpl & - & - & - & - & $1.0 \cdot 10^{-5}$ & $1.2 \cdot 10^{1}$ \\
	170405777 & 219.8 & -25.2 & $218.6 \pm 2.2$ & $-23.5 \pm 2.4$ & 2.0 & cpl & - & - & - & - & $7.4 \cdot 10^{-5}$ & $1.6 \cdot 10^{1}$ \\
	170607971 & 7.4 & 9.2 & $5.4 \pm 1.2$ & $12.5 \pm 1.1$ & 3.8 & pl & 10.8 & 8.0 & 1.9 & 3.6 & $9.4 \cdot 10^{-6}$ & $1.5 \cdot 10^{1}$ \\
	170626401 & 165.4 & 56.5 & $160.6 \pm 1.6$ & $56.1 \pm 0.7$ & 2.7 & cpl & - & - & - & - & $1.5 \cdot 10^{-5}$ & $3.7 \cdot 10^{1}$ \\
	170705115 & 191.7 & 18.3 & $192.2 \pm 1.0$ & $17.9 \pm 0.5$ & 0.7 & cpl & 187.1 & 21.2 & 2.6 & 5.2 & $1.3 \cdot 10^{-5}$ & $2.2 \cdot 10^{1}$ \\
	170906030 & 203.9 & -47.1 & $201.4 \pm 0.6$ & $-45.0 \pm 0.5$ & 2.8 & cpl & 200.6 & -46.0 & 2.1 & 2.5 & $9.5 \cdot 10^{-5}$ & $2.2 \cdot 10^{1}$ \\
	170906039 & 232.2 & -28.3 & $233.5 \pm 1.7$ & $-29.9 \pm 1.6$ & 2.0 & cpl & 235.4 & -33.6 & 3.7 & 6.0 & $3.0 \cdot 10^{-6}$ & $1.0 \cdot 10^{1}$ \\
	171120556 & 163.8 & 22.5 & $164.5 \pm 0.3$ & $23.5 \pm 0.4$ & 1.3 & cpl & - & - & - & - & $1.6 \cdot 10^{-5}$ & $3.6 \cdot 10^{1}$ \\
	180113116 & 19.2 & 68.7 & $17.0 \pm 2.7$ & $63.3 \pm 1.4$ & 5.4 & cpl & 17.7 & 66.3 & 1.0 & 2.4 & $1.4 \cdot 10^{-5}$ & 9.3 \\
	180404091 & 53.4 & -50.2 & $56.6 \pm 1.3$ & $-52.3 \pm 0.8$ & 2.9 & cpl & 53.0 & -49.3 & 1.2 & 0.9 & $2.8 \cdot 10^{-5}$ & 8.0 \\
	\enddata
\end{deluxetable*}

\newpage\clearpage

\renewcommand\refname{References}
\setlength{\itemsep}{-0.1ex plus0.1ex}
\bibliography{improved_locations}

\end{document}